\documentclass[11pt,oneside,a4paper]{article}
\usepackage[left=1in,right=1in,top=1in,bottom=1in]{geometry}
\usepackage{amsmath,amssymb}

\usepackage{hyperref}

\usepackage{graphics}
\usepackage{graphicx}
\usepackage{caption}
\usepackage{subcaption}

\usepackage{setspace}

\hypersetup{colorlinks=true, linkcolor=blue, citecolor=red, bookmarks=false, pdfstartview={FitH}}

\begin{document}

\title{\bf Notes on f(R) Theories of Gravity}

\author{
{\bf Ciprian A. Sporea}\\
\ \ \\
{\small West University of Timi\c soara, Faculty of Physics,}\\
{\small V. Parvan Ave. no. 4, 300223, Timi\c soara, Romania} \\
{\small e-mail 1: ciprian.sporea89@e-uvt.ro} \\
{\small e-mail 2: sporea\_89@yahoo.com}
}

\date{\small \today}

\maketitle

\begin{abstract}
In this review paper we present some basic notions about f(R) theories of gravity and some simple cosmological models derived from it. We first make an introduction to General Relativity (GR), followed by the discussion of Gibbons-York-Hawking boundary term in GR. We also discuss boundary terms in f(R) theories and the application of conformal transformations in order to show that f(R) theories can be made equivalent to GR minimally coupled with a scalar filed. In the final sections of the paper a brief review of classical Friedman and Lemaitre cosmological models is made, followed by the discussion of cosmological models derived from f(R)gravity.
\ \ \\


{\bf PACS (2008):} 98.80.-k $\bullet$ 04.50.Kd \\

\end{abstract}

\pagenumbering{roman}

\newpage

\tableofcontents

\newpage

\pagenumbering{arabic}

\section{Introduction}

    General Relativity (GR) continues to be, even after almost one hundred years, our best theory of gravity. Using GR we are able to derive simple cosmological models, such as the Friedman or Lemaitre models, that describe quite well the evolution of the Universe in which we live in. Up to now GR has passed every experimental test that we were able to come up with. In the last decades in order to explain the astrophysical observations related to rotation curves of spiral galaxies we where more or less forced to introduce the concept of dark matter. It did not pass long time and once again we where forced to introduce the so called dark energy in order to explain the accelerated expansion of the Universe suggested by astrophysical observation of supernovae redshift.

    Over time there have been various attempts to modify GR having as motivation different reasons, like the desire to quantize GR in order to be able to unify it with the other three elementary known forces: electromagnetic, weak and strong nuclear forces. One of the most simple modifications that can be made to GR is to add higher order invariants to the standard Einstein-Hilbert action, leading to the so called higher-order theories of gravity. A class of this theories is the theory known under the name of $f(R)$ gravity which is obtained from GR by adding higher powers of the Ricci scalar to the standard GR action. Among the motivations for the study of $f(R)$ theories is the fact that by adding suitable extra terms to the action we can recover the evolution of the Universe as observed today without the need of dark matter and dark energy.

    This paper is intended to present some basic notions about $f(R)$ theories of gravity and some simple cosmological models that can be derived from it. We start by making an extended review of the least action principle for GR and $f(R)$ theories of gravity. There are several formulations of $f(R)$ theories known under the name of metric formalism, Palatini formalism and metric-affine formalism, the last one being the most general. In this paper we will work only in the first two formalisms. A special attention is given to the surface boundary term (the analogue of Gibbons-York-Hawking boundary term from GR) which is treated in section 5. After that we show how $f(R)$ theories can be made equivalent to gravity with a minimally coupled scalar field with the help of conformal transformations.
    The last sections of the paper are reserved for the study of cosmological models. We begin with the famous classical GR Friedman and Lemaitre models and then continue with cosmological models inspired from $f(R)$ theories.

    Let us end with the remark that this paper is addressed first of all to master and PhD students and for this reason we give as much detailed calculations as possible and we also tried to write it in a pedagogical way. It can also be of use to physicists that are specialized in other fields than theoretical physics and GR in particular.

\section{Einstein-Hilbert action}

    One of the most powerful tool that theoretical physics posses is represented by the principle of least action. Using this form of a variational principle we can derive in an elegant way the equations characteristic for a specific phenomena. In what follows we will use this technique to obtain the Einstein equations of gravity, and later the equations for more general theories such as $ f(R) $ theories of gravity.

    We know that the most simple non-trivial scalar that can be constructed using only the metric tensor and it's derivatives is the Ricci curvature scalar $ R $. Then the most simple action for general relativity in vacuo turns out to be the Einstein-Hilbert action
    \begin{equation}\label{1}
    \mathcal{S}_{EH} =\int \! \mathrm{d^4}x\ \! \sqrt{g} \frac{1}{2\kappa}R
    \end{equation}
    If we take into account also the presence of matter fields, then the general action can be written as
    \begin{equation}\label{2}
    \mathcal{S} =\int \! \mathrm{d^4}x\ \! \sqrt{g}\left( \frac{1}{2\kappa}L_{EH}+L_{M} \right)
    \end{equation}
    where $ g=|\det(g_{\mu\nu})| $ and the Einstein-Hilbert and the matter lagrangians are given by
    \begin{equation}\label{3}
    L_{EH}=R  \ \ \ \ \ \ \ \ \ L_{M}=L(\psi,\partial_{\mu}\psi;g_{\mu\nu})
    \end{equation}
    In (\ref{3}) we denoted by $\psi $ the matter fields and $\kappa=8\pi G$, where $ G $ is the Newton's gravitational constant (and the speed of light is set to one by choosing the natural system of units $c=\hbar=1$).

    Now we will give a detailed review for the derivation of Einstain equations following \cite{1}, \cite{2} and \cite{3}.
    In order to obtain GR equations from action (\ref{2}) one needs to calculate the variation of the action with respect to the metric tensor.

    We consider a variation for the metric tensor of the form
    \begin{equation}\label{4}
    g_{\mu\nu}\rightarrow g_{\mu\nu}+\delta g_{\mu\nu}
    \end{equation}
    For now we will assume that both the variation of the metric ant it's first derivatives are vanishing on the boundary $ \partial\,\Sigma $ (where $ \Sigma $ is a hipervolume in the space-time manifold)
    \begin{equation}\label{5}
    \delta g_{\mu\nu}\bigg|_{\partial\,\Sigma}=0, \ \ \ \ \ \ \ \ \delta(\partial_{\mu}g_{\mu\nu})\bigg|_{\partial\,\Sigma}=0
    \end{equation}
    Let us now write the first-order variation of the Einstein-Hilbert action (\ref{2}) as \cite{1}
    \begin{equation}\label{6}
    \delta\mathcal{S}_{EH}=\frac{1}{2\kappa}\int \! \mathrm{d^4}x\ \! \left( \delta(\sqrt{g})g^{\mu\nu}R_{\mu\nu} + \sqrt{g}\delta(g^{\mu\nu})R_{\mu\nu} + \sqrt{g}g^{\mu\nu}\delta(R_{\mu\nu}) \right)
    \end{equation}
    where we use the fact that the Ricci scalar can be written as
    \begin{equation}\label{7}
    R=g^{\mu\nu}R_{\mu\nu}
    \end{equation}
    In order to calculate the variation of the determinant $ g $ of the metric we will need the following formula from linear algebra
    \begin{equation}\label{8}
    g=g_{\mu\nu}adj(g_{\mu\nu})
    \end{equation}
    from which follows immediately
    \begin{equation}\label{9}
    \delta g=adj(g_{\mu\nu})\delta g_{\mu\nu}
    \end{equation}
    and taking into account that by definition the adjoint of a matrix is $ adj=g\,g^{\mu\nu} $, one gets in the end
    \begin{equation}\label{10}
    \delta g=g\,g^{\mu\nu}\delta g_{\mu\nu}
    \end{equation}
    Because we are choosing to make the variation of the action with respect to $ \delta g^{\mu\nu} $ instead of $ \delta g_{\mu\nu} $ (although in the end the same equations are obtained in both cases) we will derive now a relation between $ \delta g^{\mu\nu} $ and $ \delta g_{\mu\nu} $. In order to do that we will start from the following relation
    \begin{equation}\label{11}
    g^{\mu\sigma}g_{\sigma\nu}=\delta^\mu_\nu
    \end{equation}
    and using the fact that the constant tensor $ \delta^\mu_\nu $ dose not change under a variation, one may therefore write
    \begin{equation}\label{12}
    \delta g^{\mu\sigma}g_{\sigma\nu}+g^{\mu\sigma}\delta g_{\sigma\nu}=0
    \end{equation}
    or after multiplying with $ g^{\rho\nu} $ and rearranging it, one gets
    \begin{equation}\label{13}
    \delta g_{\mu\rho}=-g^{\mu\sigma}g^{\rho\nu}\delta g_{\sigma\nu}
    \end{equation}
    We now relabel the indices $ \rho \leftrightarrow \nu $ and multiply again by $ g_{\mu\nu} $ to get
    \begin{equation}\label{14}
    \begin{split}
    g_{\mu\nu}\delta g^{\mu\nu}&=-g_{\mu\nu}g^{\mu\sigma}g^{\rho\nu}\delta g_{\sigma\rho}=-\delta^\sigma_\nu g^{\rho\nu}\delta g_{\sigma\rho}\\
    &=-g^{\sigma\rho}\delta g_{\sigma\rho}=-g^{\mu\nu}\delta g_{\mu\nu}
    \end{split}
    \end{equation}
    Inserting the above expression in (\ref{10}) we can now calculate the variation for $ \sqrt{g} $ \cite{3} as
    \begin{equation}\label{15}
    \begin{split}
    \delta\sqrt{g}&=\frac{1}{2\sqrt{g}}\delta g=\frac{-1}{2\sqrt{g}}gg_{\mu\nu}\delta g^{\mu\nu}\\
    &=-\frac{1}{2}\sqrt{g}g_{\mu\nu}\delta g^{\mu\nu}
    \end{split}
    \end{equation}
    In order to extract the field equations from (\ref{6}) we need to factor out the variation $ \delta g^{\mu\nu} $. To do this we first need to express $ \delta R_{\mu\nu} $ in terms of the variation of the metric and it's derivatives. Because the Ricci tensor is obtained from the full Riemann curvature tensor by
    \begin{equation}\label{16}
    R_{\mu\nu}=R^\sigma_{\ \mu\sigma\nu}
    \end{equation}
    in what follows we will derive the variation of the Riemann curvature tensor. This can be done in a number of ways, but the most straightforward one is to calculate first the variation of the curvature tensor. The Riemann tensor is given by \cite{2}
    \begin{equation}\label{17}
    R^\sigma_{\ \mu\nu\rho} = \partial_\nu\Gamma^\sigma_{\ \mu\rho}-\partial_\rho\Gamma^\sigma_{\ \mu\nu} + \Gamma^\lambda_{\ \mu\rho}\Gamma^\sigma_{\ \lambda\nu}-\Gamma^\lambda_{\ \mu\nu}\Gamma^\sigma_{\ \lambda\rho}
    \end{equation}
    in terms of the connections $  \Gamma^\sigma_{\ \mu\nu} $. For GR the standard connection is the Levi-Civita one
    \begin{equation}\label{18}
    \Gamma^\sigma_{\ \mu\nu}=\frac{1}{2}g^{\sigma\lambda}\left( \partial_\mu g_{\lambda\nu}+\partial_\nu g_{\mu\lambda}-\partial_\lambda g_{\mu\nu} \right)
    \end{equation}
    Under an arbitrary variation of the connection coefficients
    \begin{equation}\label{18a}
    \Gamma^\sigma_{\ \mu\nu} \rightarrow \Gamma^\sigma_{\ \mu\nu}+\delta\Gamma^\sigma_{\ \mu\nu}
    \end{equation}
    we can write from (\ref{17}) that
    \begin{equation}\label{19}
    \delta R^\sigma_{\ \mu\nu\rho} = \partial_\nu(\delta\Gamma^\sigma_{\ \mu\rho})-\partial_\rho(\delta\Gamma^\sigma_{\ \mu\nu}) + \delta\Gamma^\lambda_{\ \mu\rho}\Gamma^\sigma_{\ \lambda\nu}-\delta\Gamma^\lambda_{\ \mu\nu}\Gamma^\sigma_{\ \lambda\rho} + \Gamma^\lambda_{\ \mu\rho}\delta\Gamma^\sigma_{\ \lambda\nu}-\Gamma^\lambda_{\ \mu\nu}\delta\Gamma^\sigma_{\ \lambda\rho}
    \end{equation}
    Do to the fact that we can choose to work in local geodesic coordinates at some arbitrary point $ P $ at which the Christofel symbols can be made zero ($ \Gamma^\alpha_{\ \sigma\mu}=0 $) \cite{34} and in that point also holds that the ordinary derivative is identical with the covariant derivative defined by
    \begin{equation}\label{20}
    \nabla_\mu T\,^\alpha_{\ \!\beta} = \partial_\mu T\,^\alpha_{\ \!\beta} + \Gamma^\alpha_{\ \sigma\mu}T\,^\sigma_{\ \!\beta} - \Gamma^\sigma_{\ \beta\mu}T\,^\alpha_{\ \!\sigma}
    \end{equation}
    we can write the variation of the Riemann tensor at the point $ P $ as
    \begin{equation}\label{21}
    \delta R^\sigma_{\ \mu\nu\rho} = \nabla_\nu(\delta\Gamma^\sigma_{\ \mu\rho})-\nabla_\rho\delta(\Gamma^\sigma_{\ \mu\nu})
    \end{equation}
    The fact that $ \delta\Gamma^\sigma_{\ \mu\rho} $ is a tensor and the RHS of (\ref{21}) contains only tensorial quantities tells us that this relation holds in any coordinate system (see \cite{3} or \cite{34} for more details). The result (\ref{21}) thus holds generally and it is known as the Palatini equation.
    Using (\ref{16}) and (\ref{21}) we can now write the variation for the Ricci tensor
    \begin{equation}\label{22}
    \delta R_{\mu\nu} = \nabla_\nu(\delta\Gamma^\sigma_{\ \mu\sigma})-\nabla_\sigma\delta(\Gamma^\sigma_{\ \mu\nu})
    \end{equation}
    It now follows that
    \begin{equation}\label{23}
    \begin{split}
    g^{\mu\nu}\delta R_{\mu\nu} &= g^{\mu\nu}\nabla_\nu(\delta\Gamma^\sigma_{\ \mu\sigma})- g^{\mu\nu}\nabla_\sigma\delta(\Gamma^\sigma_{\ \mu\nu})\\
    &=\nabla_\nu(g^{\mu\nu}\delta\Gamma^\sigma_{\ \mu\sigma})- \nabla_\nu(g^{\mu\sigma}\delta\Gamma^\nu_{\ \mu\sigma})\\
    &=\nabla_\nu(g^{\mu\nu}\delta\Gamma^\sigma_{\ \mu\sigma}-g^{\mu\sigma}\delta\Gamma^\nu_{\ \mu\sigma})
    \end{split}
    \end{equation}
    Introducing this together with relation (\ref{15}) into (\ref{6}) we have the following expression for the variation of the action (\ref{1})
    \begin{equation}\label{24}
    \begin{split}
    \delta\mathcal{S}_{EH} =&\int\!\mathrm{d^4}x\,\sqrt{g} \left\{-\frac{1}{2}g_{\mu\nu}R\right\}\delta g^{\mu\nu} + \int\!\mathrm{d^4}x\,\sqrt{g}R_{\mu\nu}\delta g^{\mu\nu} + \\
    &+ \int\!\mathrm{d^4}x\,\sqrt{g}\,\nabla_\nu\left( g^{\mu\nu}\delta\Gamma^\sigma_{\ \mu\sigma}-g^{\mu\sigma}\delta\Gamma^\nu_{\ \mu\sigma}\right)
    \end{split}
    \end{equation}
    The last integral of relation (\ref{24}) can be transformed into a surface integral via the Gauss-Stokes theorem and because on the surface we impose the conditions (\ref{5}) then the surface integral will vanish. However, this will no longer be the case for the $ f(R) $ action when the surface term can not be made zero in general (see \cite{43}, \cite{4}, \cite{5} for more details). Taking this into account the final expression for the variation of $ \mathcal{S_{EH}} $ can be written as
    \begin{equation}\label{25}
    \delta\mathcal{S}_{EH} =\int\!\mathrm{d^4}x\,\sqrt{g} \left\{R_{\mu\nu}-\frac{1}{2}g_{\mu\nu}R\right\}\delta g^{\mu\nu}
    \end{equation}
    After calculating the variational integral $ \delta\mathcal{S}_{EH}/\delta g^{\mu\nu} $ and setting it to zero we obtain in the end the Einstein equations of gravity for empty space
    \begin{equation}\label{26}
    R_{\mu\nu}-\frac{1}{2}g_{\mu\nu}R=0
    \end{equation}
    For the remaining of this section we will analyse the full action (\ref{2}), which can also be written as \cite{1}
    \begin{equation}\label{27}
    \mathcal{S}=\frac{1}{2\kappa}\mathcal{S}_{EH}+\mathcal{S}_{M}=\int\!\mathrm{d^4}x\,\left( \frac{1}{2\kappa}\mathcal{L}_{EH}+\mathcal{L}_{M} \right)
    \end{equation}
    where we denoted with $ \mathcal{L}_{EH} $ and $ \mathcal{L}_{M} $ the lagrange densities given by
    \begin{equation}\label{28}
    \mathcal{L}_{EH}=\sqrt{g}L_{EH}  \ \ \ \ \ \ \ \ \ \ \mathcal{L}_{M}=\sqrt{g}L_{M}
    \end{equation}
    Varying action (\ref{27}) with respect to (inverse) metric we obtain
    \begin{equation}\label{29}
    \frac{1}{2\kappa}\frac{\delta\mathcal{S}_{EH}}{\delta g^{\mu\nu}}+\frac{\delta\mathcal{S}_{M}}{\delta g^{\mu\nu}}=0
    \end{equation}
    We already saw that the first term of the above relation gives us the Einstein tensor $ G_{\mu\nu} $ defined by (\ref{25}) or (\ref{26}) as
    \begin{equation}\label{30}
    G_{\mu\nu}=R_{\mu\nu}-\frac{1}{2}g_{\mu\nu}R
    \end{equation}
    Comparing (\ref{29}) with the full Einstein equations
    \begin{equation}\label{31}
    G_{\mu\nu}=\kappa T_{\mu\nu}
    \end{equation}
    we can now give a definition for the stress-energy momentum tensor as a function of the metric and the matter lagrangian
    \begin{equation}\label{32}
    T_{\mu\nu}=\frac{-2}{\sqrt{g}}\frac{\delta\mathcal{S}_{M}}{\delta g^{\mu\nu}}
    \end{equation}
    It can be shown (see for example \cite{1}) that the stress-energy momentum tensor (\ref{32}) has all the proprieties required of an energy-momentum tensor.

\section{Gibbons-York-Hawking boundary term}

    In the previous section we've argued that the last integral in relation (\ref{24}) will vanish if we transform this integral via the Gauss-Stokes theorem and we impose the supplementary condition $ \delta(\partial_{\mu}g_{\mu\nu})\bigg|_{\partial\,\Sigma}=0 $. Let us now see what happens when we drop out this condition. Then on the boundary the variation of the metric derivatives will no longer vanish, case in which we will have a non-vanishing boundary term. If this is the case, then from the variation of the action (\ref{1}) given by (\ref{24}) we will no longer obtain the Einstein GR equations. In order to fix this problem, one needs to amend the action (\ref{1}) with a new term, called the Gibbons-York-Hawking boundary term \cite{6},\cite{7}, \cite{30} (which can be seen as a contraterm)
    \begin{equation}\label{33}
    \mathcal{S}=\mathcal{S}_{EH}+\mathcal{S}_{B}
    \end{equation}
    The variation of this new boundary term is (conform (\ref{24}))
    \begin{equation}\label{34}
    \delta\mathcal{S}_{B}=\int\!\mathrm{d^4}x\,\sqrt{g}\,\nabla_\nu\left( g^{\mu\nu}\delta\Gamma^\sigma_{\ \mu\sigma}-g^{\mu\sigma}\delta\Gamma^\nu_{\ \mu\sigma}\right)
    \end{equation}
    and if we define
    \begin{equation}\label{34a}
    V^\nu=g^{\mu\nu}\delta\Gamma^\sigma_{\ \mu\sigma}-g^{\mu\sigma}\delta\Gamma^\nu_{\ \mu\sigma}
    \end{equation}
    then the boundary term can be written as
    \begin{equation}\label{34b}
    \delta\mathcal{S}_{B}=\int\!\mathrm{d^4}x\,\sqrt{g}\,\nabla_\nu V^\nu
    \end{equation}
    In order to express this variation in terms of $ \delta g^{\mu\nu} $ we first need to compute the variation of the connection coefficients. Let us rewrite here the Levi-Civita connection (\ref{18})
    \begin{equation}\label{35}
    \Gamma^\sigma_{\ \mu\nu}=\frac{1}{2}g^{\sigma\lambda}\left( \partial_\mu g_{\lambda\nu}+\partial_\nu g_{\mu\lambda}-\partial_\lambda g_{\mu\nu} \right)
    \end{equation}
    From this follows immediately
    \begin{equation}\label{36}
    \begin{split}
    \delta\Gamma^\sigma_{\ \mu\nu} &= \delta\left\{\frac{1}{2}g^{\sigma\lambda}\left( \partial_\mu g_{\lambda\nu}+\partial_\nu g_{\mu\lambda}-\partial_\lambda g_{\mu\nu} \right)\right\}\\
    &=\frac{1}{2}\delta g^{\sigma\lambda}\left( \partial_\mu g_{\lambda\nu}+\partial_\nu g_{\mu\lambda}-\partial_\lambda g_{\mu\nu} \right) + \frac{1}{2}g^{\sigma\lambda}\left\{ \partial_\mu(\delta g_{\lambda\nu})+\partial_\nu (\delta g_{\mu\lambda})-\partial_\lambda (\delta g_{\mu\nu}) \right\}
    \end{split}
    \end{equation}
    Taking into account the boundary condition $ \delta g_{\mu\nu}=\delta g^{\mu\nu}=0 $ the variation (\ref{36}) gives
    \begin{equation}\label{37}
    \delta\Gamma^\sigma_{\ \mu\nu}\bigg|_{\partial\,\Sigma}=\frac{1}{2}g^{\sigma\lambda}\left\{ \partial_\mu(\delta g_{\lambda\nu})+\partial_\nu (\delta g_{\mu\lambda})-\partial_\lambda (\delta g_{\mu\nu}) \right\}
    \end{equation}
    Using this result for computing $ V^\nu $ in (\ref{34a}), we have
    \begin{equation}\label{38}
    \begin{split}
    V^\nu\bigg|_{\partial\,\Sigma}&=\frac{1}{2}g^{\mu\nu}g^{\sigma\lambda}\left\{ \partial_\mu(\delta g_{\lambda\sigma})+\partial_\sigma (\delta g_{\mu\lambda})-\partial_\lambda (\delta g_{\mu\sigma}) \right\} \\
    &-\frac{1}{2}g^{\mu\sigma}g^{\nu\lambda}\left\{ \partial_\mu(\delta g_{\lambda\sigma})+\partial_\sigma (\delta g_{\mu\lambda})-\partial_\lambda (\delta g_{\mu\sigma}) \right\}\\
    &=\frac{1}{2}g^{\mu\nu}g^{\sigma\lambda} \partial_\mu(\delta g_{\lambda\sigma}) - \frac{1}{2}g^{\mu\sigma}g^{\nu\lambda}\left\{ \partial_\mu(\delta g_{\lambda\sigma})+\partial_\sigma (\delta g_{\mu\lambda})-\partial_\lambda (\delta g_{\mu\sigma}) \right\}
    \end{split}
    \end{equation}
    We can transform (\ref{34b}) by applying the Gauss-Stokes theorem \cite{2},\cite{8}
    \begin{equation}\label{39}
    \int\!\mathrm{d^n}x\,\sqrt{g}\,\nabla_\mu A^\mu = \oint_{\partial\Sigma}\!\mathrm{d^{n-1}}y\,\sqrt{|h|}\,\epsilon\,n_\mu A^\mu
    \end{equation}
    where $h$ is the determinant of the induced metric, $ n_\mu $ is the unit normal 4-vector to $ \partial\Sigma $ and $\epsilon=n_\mu n^\mu=\pm1 $ if $\partial\Sigma$ is a timelike hipersurface or a spacelike one.
    Now we can write $\delta\mathcal{S}_{B}$ as
    \begin{equation}\label{40}
    \begin{split}
    \delta\mathcal{S}_{B}&=\oint_{\partial\Sigma}\!\mathrm{d^{3}}y\,\sqrt{|h|}\,\epsilon\,n_\nu V^\nu\\
    &=\oint_{\partial\Sigma}\!\mathrm{d^{3}}y\,\sqrt{|h|}\,\epsilon\,n^\nu V_\nu
    \end{split}
    \end{equation}
    Let us now compute the quantity $ n^\nu V_\nu $. Using (\ref{38}) we get
    \begin{equation}\label{41}
    \begin{split}
    n^\nu V_\nu\bigg|_{\partial\,\Sigma}&=n^\nu g_{\gamma\nu}V^\gamma\bigg|_{\partial\,\Sigma} =\frac{1}{2}n^\nu g_{\gamma\nu}g^{\mu\gamma}g^{\sigma\lambda} \partial_\mu(\delta g_{\lambda\sigma}) \\
    &-\frac{1}{2}n^\nu g_{\gamma\nu}g^{\mu\sigma}g^{\gamma\lambda}\left\{ \partial_\mu(\delta g_{\lambda\sigma})+\partial_\sigma (\delta g_{\mu\lambda})-\partial_\lambda (\delta g_{\mu\sigma}) \right\}\\
    &=\frac{1}{2}n^\nu \delta^\mu_{\ \nu}g^{\sigma\lambda} \partial_\mu(\delta g_{\lambda\sigma})- \frac{1}{2}n^\nu\delta^\lambda_{\ \nu}g^{\mu\sigma}\left\{ \partial_\mu(\delta g_{\lambda\sigma})+\partial_\sigma (\delta g_{\mu\lambda})-\partial_\lambda (\delta g_{\mu\sigma}) \right\}\\
    &=\frac{1}{2}n^\nu g^{\sigma\lambda} \partial_\nu(\delta g_{\lambda\sigma})-\frac{1}{2}n^\nu g^{\mu\sigma} \partial_\mu(\delta g_{\nu\sigma})- \frac{1}{2}n^\nu g^{\mu\sigma} \partial_\sigma(\delta g_{\mu\nu})+ \frac{1}{2}n^\nu g^{\mu\sigma} \partial_\nu(\delta g_{\mu\sigma})\\
    &=n^\nu g^{\mu\sigma}\left\{\partial_\nu(\delta g_{\mu\sigma}) -\partial_\mu(\delta g_{\nu\sigma}) \right\}
    \end{split}
    \end{equation}
    The connection between the metric $g_{\mu\nu}$ and the induced metric is given by the following relation \cite{2}
    \begin{equation}\label{42}
    g^{\alpha\beta}=h^{\alpha\beta}+\epsilon n^\alpha n^\beta
    \end{equation}
    then
    \begin{equation}\label{43}
    \begin{split}
    n^\nu V_\nu\bigg|_{\partial\,\Sigma}&=n^\nu h^{\mu\sigma}\left\{\partial_\nu(\delta g_{\mu\sigma}) -\partial_\mu(\delta g_{\nu\sigma}) \right\} + n^\nu\epsilon n^\mu n^\sigma\left\{\partial_\nu(\delta g_{\mu\sigma}) -\partial_\mu(\delta g_{\nu\sigma}) \right\}\\
    &=n^\nu h^{\mu\sigma}\left\{\partial_\nu(\delta g_{\mu\sigma}) -\partial_\mu(\delta g_{\nu\sigma}) \right\}
    \end{split}
    \end{equation}
    where in the second line we used the fact that $ n^\mu n^\sigma $ is multiplied by the antisymmetric quantity within the brackets.

    Now, because the tangential derivatives of $\delta g_{\mu\nu}$ are vanishing on the boundary $\delta\Sigma$ eq. (\ref{43}) becomes
    \begin{equation}\label{44}
    n^\nu V_\nu\bigg|_{\partial\,\Sigma}=n^\nu h^{\mu\sigma}\partial_\nu(\delta g_{\mu\sigma})
    \end{equation}
    Thus the variation of the boundary term will read
    \begin{equation}\label{45}
    \delta\mathcal{S}_{B}=\oint_{\partial\Sigma}\!\mathrm{d^{3}}y\,\sqrt{|h|}\,\epsilon\,n^\nu h^{\mu\sigma}\partial_\nu(\delta g_{\mu\sigma})
    \end{equation}
    The Gibbons-York-Hawking boundary term \cite{6},\cite{7}, \cite{30} is given by the following expresion
    \begin{equation}\label{46}
    \mathcal{S}_{GYH}=-2\oint_{\partial\Sigma}\!\mathrm{d^{3}}y\,\sqrt{|h|}\,\epsilon K
    \end{equation}
    where $K$ is the trace of the extrinsic curvature of the boundary $\partial\Sigma$. Let us now show that this formula is equivalent with the expression given by relation (\ref{45}).

    Because the induced metric $h_{\mu\nu}$ is fixed on $\partial\Sigma$ the variation of $\mathcal{S}_{GYH}$ will be
    \begin{equation}\label{47}
    \delta\mathcal{S}_{GYH}=-2\oint_{\partial\Sigma}\!\mathrm{d^{3}}y\,\sqrt{|h|}\,\epsilon\delta K
    \end{equation}
    In order to calculate the variation of the extrinsic curvature, we start by recalling it's definition \cite{2}
    \begin{equation}\label{48}
    \begin{split}
    K&=\nabla_\mu n^\mu=g^{\mu\nu}\nabla_\mu n_\nu\\
    &=(h^{\mu\nu}+\epsilon n^\mu n^\nu)\nabla_\mu n_\nu\\
    &=h^{\mu\nu}\nabla_\mu n_\nu\\
    &=h^{\mu\nu}(\partial_\mu n_\nu - \Gamma^\sigma_{\ \mu\nu}n_\sigma)
    \end{split}
    \end{equation}
    Using the above, the variation of $K$ will be
    \begin{equation}\label{49}
    \begin{split}
    \delta K&=\delta(h^{\mu\nu})(\partial_\mu n_\nu - \Gamma^\sigma_{\ \mu\nu}n_\sigma)+h^{\mu\nu}\delta(\partial_\mu n_\nu - \Gamma^\sigma_{\ \mu\nu}n_\sigma)\\
    &=-h^{\mu\nu}\delta(\Gamma^\sigma_{\ \mu\nu})n_\sigma\\
    &=-\frac{1}{2}h^{\mu\nu}g^{\sigma\lambda}\left\{ \partial_\mu(\delta g_{\lambda\nu})+\partial_\nu (\delta g_{\mu\lambda})-\partial_\lambda (\delta g_{\mu\nu}) \right\}n_\sigma\\
    &=-\frac{1}{2}h^{\mu\nu}\left\{ \partial_\mu(\delta g_{\lambda\nu})+\partial_\nu (\delta g_{\mu\lambda})-\partial_\lambda (\delta g_{\mu\nu}) \right\}n^\lambda\\
    &=\frac{1}{2}h^{\mu\nu}\partial_\lambda (\delta g_{\mu\nu})n^\lambda
    \end{split}
    \end{equation}
    Introducing this result into (\ref{47}) we obtain
    \begin{equation}\label{50}
    \delta\mathcal{S}_{GYH}=-\oint_{\partial\Sigma}\!\mathrm{d^{3}}y\,\sqrt{|h|}\,\epsilon\,n^\lambda h^{\mu\nu}\partial_\lambda(\delta g_{\mu\nu})
    \end{equation}
    which shows that by adding $\mathcal{S}_{GYH}$ to the original $\mathcal{S}_{EH}$ action we recover the Einstein field equations.

\section{f(R) theories of gravity}

\subsection{Introduction}

    General Relativity (GR) is a comprehensive theory of space-time, gravity and matter. Based on GR nowadays we have a model of the Universe called the Standard Cosmology Model. More recently, new evidence coming from astrophysics and cosmology  are showing that the Universe is currently going through a faze of accelerated expansion \cite{39},\cite{40}. It is thought that this expansion is determined by the existence of an unknown form of energy called dark energy which has not been detected directly and dose not cluster as ordinary matter dose. The latest data provided by the Plank mission of Cosmic Microwave Background Radiation (CMBR) give us the following composition for the total mass-energy content of the Universe: 4.9\% ordinary baryonic matter, 26.8\% dark matter and 68.3\% dark energy \cite{9}. The term dark matter refers to an unknown form of matter, which has the clustering properties of ordinary matter but has not yet been observed or detected in the laboratory.

    It is assumed that the Universe had also an early time accelerated epoch as predicted by the inflationary paradigm \cite{19},\cite{24},\cite{29}. The inflationary epoch is needed to address the so-called horizon, flatness and monopole problems \cite{24},\cite{29},\cite{35},\cite{8} as well as to provide the mechanism that generates primordial inhomogeneities acting as seeds for the formation of large scale structures \cite{36},\cite{37}.

    The simplest model which adequately fits the new observations described above is the concordance model or $ \Lambda CDM $ ($\Lambda$-Cold Dark Matter) supplemented by some inflationary scenario, usually based on some scalar field called inflaton. However, this model dose not explain the origin of inflation or the nature of dark matter and like other models is burdened with the well known cosmological constant problem \cite{10},\cite{53}.

    A different approach for solving the problems raised by the new cosmological observations is to consider the alternative of modifying General Relativity. In this category enter the so called higher-order theories of gravity, i.e., modifications of the Einstein-Hilbert action in order to include higher-order curvature invariants with respect to the Ricci scalar (see \cite{41} for a historical review and a list of references to early work). One can deviate from GR in various ways, the most well known alternative is the scalar-tensor Brans-Dike theory \cite{12}. Other typical examples are DGP gravity (Dvali-Gabadadze-Porrati) \cite{13}, braneworld gravity \cite{32}, TeVeS (Tensor-Vector-Scalar) \cite{11} and Einstein-Aether theory \cite{23}.

    In what follows we will describe one class of modified theories of gravity known as f(R) theories of gravity (for reviews on f(R) one can consult \cite{43}, \cite{77}, \cite{78}). These theories are obtained by modifying the Einstein-Hilbert action
    \begin{equation}\label{3.1}
    \mathcal{S}_{EH} =\int\!\mathrm{d^4}x\,\sqrt{-g} \left \{\frac{1}{2}R \right \}
    \end{equation}
    in the following way
    \begin{equation}\label{3.2}
    \mathcal{S} =\int\!\mathrm{d^4}x\,\sqrt{-g}f(R)
    \end{equation}
    where $f(R)$ is an arbitrary function of the Ricci scalar $ R $.
    The appealing feature of this action is that it combines mathematical simplicity with a fair amount of generality. For instance if we take a series expansion of $ f $
    \begin{equation}\label{3.3}
    f(R) = ...+\frac{a_2}{R^2}+\frac{a_1}{R}-2\Lambda+R+ b_2R^2+b_3R^3+...
    \end{equation}
    where the $ a_i $ and $ b_j $ coefficients have the appropriate dimensions, we see that the action includes a number of phenomenologically interesting terms. $ f(R) $ actions where first rigourously studied by Buchdahl \cite{76}.

    Higher order action can include also other curvature invariants, such as $ R^2 $, $ R_{\mu\nu}R^{\mu\nu} $, $ \Box R $ etc. An interesting example is the Brans-Dicke action
    \begin{equation}\label{3.4}
    \mathcal{S}_{BD} =\int\!\mathrm{d^4}x\,\sqrt{-g} \left \{ \phi R - \frac{\omega_0}{\phi}(\partial_\mu\phi\,\partial^{\,\mu}\phi) \right \} + \mathcal{S}_M(g_{\mu\nu},\psi)
    \end{equation}
    where $ \phi $ is a scalar field and $ \omega_0 $ is called the Brans-Dicke parameter.

    It can be shown that f(R) gravity is equivalent with the Brans-Dicke theory for $ \omega_0=0 $ in the metric formalism and $ \omega_0=-3/2 $ in the Palatini formulation of f(R) theory (see reference \cite{42}).

    In order to obtain the Einstein equations we apply a variation principle on action (\ref{3.2}). This can be done in two ways depending on which variational principle one uses \cite{1},\cite{34},\cite{8}. The standard approach is the variation of action $ \mathcal{S} $ with respect to the metric tensor $ g^{\mu\nu} $ (known also as the metric formalism), while the less standard variation, called Palatini variation, is when one varies the action with respect to the connection and the metric if we are assuming that the two are independent variables. In the case of Palatini formalism we are also making the assumption that the matter action dose not depend on the connection. We will show in this section that the two formalisms give us in general different field equations with the observation that for a linear action in $ R $ we obtain in both cases the usual Einstein equation (\ref{31}). Thus we see that we have two versions of $f(R)$ gravity: f(R) gravity obtained in the metric formalism and Palatini $f(R)$ gravity if we choose to work with the Palatini variation. If we give up to the assumption that the matter action is independent of the connection we end up with a third version of $ f(R) $ gravity, known under the name of {\it metric-affine $ f(R) $ gravity} \cite{44},\cite{45}, which is the most general of these theories.

\subsection{Metric formalism}

    The action in the metric formalism is made of two terms
    \begin{equation}\label{3.5}
    \mathcal{S}_{met} = \mathcal{S}_G(g_{\mu\nu}) + \mathcal{S}_M(g_{\mu\nu},\psi)
    \end{equation}
    where $ \mathcal{S}_M $ is the matter action and we denoted by $ \psi $ the matter fields.

    Thus we can write the total action for $ f(R) $ gravity as
    \begin{equation}\label{3.6}
    \mathcal{S}_{met} = \int\!\mathrm{d^4}x\,\sqrt{g}f(R) + \mathcal{S}_M(g_{\mu\nu},\psi)
    \end{equation}
    Varying action (\ref{3.6}) with respect to the metric, after some manipulations (see section \ref{1.4}), we obtain
    \begin{equation}\label{3.7}
    \begin{split}
    \delta\mathcal{S}_{met} &= \int\!\mathrm{d^4}x\,\sqrt{g}\, \bigg\{ f'(R)R_{\mu\nu} -\frac{1}{2}f(R)g_{\mu\nu}- \\
    &-\nabla_\mu\nabla_\nu f'(R)+g_{\mu\nu}\Box f'(R)-\kappa T_{\mu\nu}  \bigg\}\,\delta g^{\mu\nu}
    \end{split}
    \end{equation}
    where $'\equiv d/dR $, $ \nabla_\mu $ is the standard covariant derivative (obtained via Levi-Civita connection), and $ \Box=\nabla^\mu\nabla_\mu $ is the Laplace operator in four dimensions.

    From (\ref{3.7}) we can immediately read the new Einstein equations
    \begin{equation}\label{3.8}
     f'(R)R_{\mu\nu} -\frac{1}{2}f(R)g_{\mu\nu}-\left [ \nabla_\mu\nabla_\nu-g_{\mu\nu}\Box \right ] f'(R)=\kappa T_{\mu\nu}
     \end{equation}
    One immediately sees that eqs. (\ref{3.8}) are fourth order partial differential equations in the metric $ g_{\mu\nu} $. In the particular case when $ f'(R) $ is a constant we see that the forth order terms (the last two on the LHS of \ref{3.8}) will vanish. If $ f'(R) $ is a constant then obviously $ f(R) $ is a linear function of $ R $ and the theory reduces to standard General Relativity (GR).

    The trace of eqs (\ref{3.8}) is given by
    \begin{equation}\label{3.9}
    f'(R)R - 2f(R) + 3\Box f'(R) = \kappa T
    \end{equation}
    This is a differentially relation between $R$ and $T$, opposed to the algebraic one of GR, where $ R + T = 0 $. Eq. (\ref{3.9}) tells us that $f(R)$ theories can have more solutions than Einstein's classical theory.

    Eq. (\ref{3.9}) with $ T=0 $ dose not necessarily imply that $ R=0 $, or that $R$ is constant. From GR we know that a constant Ricci scalar leads us to maximally symmetric solutions. In the case of $f(R)$, for $ R=const $ and $ T_{\mu\nu}=0 $, eq. (\ref{3.9}) reduces to
     \begin{equation}\label{3.10}
    f'(R)R - 2f(R) = 0
    \end{equation}
    The form of the function $f$ will tell us the type of the maximally symmetric solutions, as follows: if $f$ is chosen such that $ R=0 $ is a root of eq. (\ref{3.10}) and we introduce this solution in eq. (\ref{3.8}), then we obtain $ R_{\mu\nu}=0 $ and the maximally symmetric space will be Minkowski spacetime; when the form of $f$ leads to the root $ R = \pm C $ for eq. (\ref{3.10}) it follows that $ R_{\mu\nu}=g_{\mu\nu}C/4 $, case in which the maximally symmetric solutions will be de Sitter ($+C$) or anti-de Sitter ($-C$) spacetime, analogue to general relativity when a cosmological constant is present.

    Let us end this subsection by rewriting the field equations (\ref{3.8}) of $ f(R) $ gravity in the form of Einstein equations as
    \begin{equation}\label{3.11}
    \begin{split}
    G_{\mu\nu} \equiv R_{\mu\nu}-\frac{1}{2}g_{\mu\nu}R =& \frac{\kappa T_{\mu\nu}}{f'(R)} + g_{\mu\nu}\frac{f(R)-Rf'(R)}{2f'(R)} \, + \\
    &+ \, \frac{\nabla_\mu\nabla_\nu f'(R)-g_{\mu\nu}\Box f'(R)}{f'(R)} \\
    &=\frac{\kappa}{f'(R)}\left[\, T_{\mu\nu} + T_{\mu\nu}^{(eff)}\, \right]
    \end{split}
    \end{equation}
    where we've introduced an effective stress-energy tensor
    \begin{equation}\label{3.13}
    T_{\mu\nu}^{(eff)}\equiv \left [ \frac{f(R)-Rf'(R)}{2}g_{\mu\nu}+\nabla_\mu\nabla_\nu f'(R)-g_{\mu\nu}\Box f'(R) \right]
    \end{equation}

\subsection{Palatini formalism}

    In what follows we will discuss the other variational principle used in gravity, namely the Palatini formalism \cite{43}, i.e. an independent variation with respect to the metric and an independent connection. The action will have the same form as in the case of metric variation. However, the Riemann tensor (denoted by $ \mathcal{R}_{\mu\nu} $) and the Ricci tensor ($ \mathcal{R} = g^{\mu\nu}\mathcal{R}_{\mu\nu} $) will not depend directly on the metric because we use for their construction the independent connection. Thus the action takes the following form
    \begin{equation}\label{3.14}
    \mathcal{S}_{pal} =\int \! \mathrm{d^4}x\ \! \sqrt{g}f(\mathcal{R}) + \mathcal{S}_M(g_{\mu\nu},\psi)
    \end{equation}
    From (\ref{3.14}) we can see that the matter action dose not depend on the independent connection, this being an important property of the Palatini $f(R)$ gravity and will also play an important role in obtaining Einstein gravity in the case of an linear function for $f(R)$.   

    We return now to derive the field equations for Palatini $ f(R) $ gravity. Due to the fact that $ \mathcal{R}_{\mu\nu} $ dose not depend on the metric, the variation of action (\ref{3.14}) with respect to $ g^{\mu\nu} $ is straightforward. However, the variation with respect to  the connection is a little more tricky because we need first to calculate the variation of $ \delta\,\mathcal{R}_{\mu\nu} $ (see section 5 for a derivation)
    \begin{equation}\label{3.15}
    \delta\,\mathcal{R}_{\mu\nu} = \bar{\nabla}_\lambda\delta\Gamma^\lambda_{\ \mu\nu} - \bar{\nabla}_\nu\delta\Gamma^\lambda_{\ \mu\lambda}
    \end{equation}
    where we denoted by $ \bar{\nabla}_\lambda $ the covariant derivative defined with the independent connection. We note that the variation of the matter action with respect to the independent connection is zero, since the mater action dose not depend on $ \Gamma^\lambda_{\ \mu\nu} $ and by definition $ T_{\mu\nu} $ is defined with the help of relation (\ref{32}).

    Using eq. (\ref{3.15}), the variation of the gravitational part of the action (\ref{3.14}) turns out to be
    \begin{equation}\label{3.16}
    \begin{split}
    \delta\mathcal{S}_{pal} &=\int \! \mathrm{d^4}x\ \! \sqrt{g}\left\{ f'(\mathcal{R})\mathcal{R}_{(\mu\nu)} - \frac{1}{2}f(\mathcal{R})g_{\mu\nu} - \kappa\,T_{\mu\nu} \right\}\delta g^{\mu\nu} + \\
    &+\int \! \mathrm{d^4}x\ \! \sqrt{g}f'(\mathcal{R})g^{\mu\nu}\left( \bar{\nabla}_\lambda\delta\Gamma^\lambda_{\ \mu\nu} - \bar{\nabla}_\nu\delta\Gamma^\lambda_{\ \mu\lambda} \right)
    \end{split}
    \end{equation}
    Integrating the last term by parts and taking into account that on the boundary $ \delta\Gamma^\lambda_{\ \mu\nu}=0 $, we get
    \begin{equation}\label{3.17}
    \begin{gathered}
    \int \! \mathrm{d^4}x\ \! \sqrt{-g}f'(\mathcal{R})g^{\mu\nu}\left( \bar{\nabla}_\lambda\delta\Gamma^\lambda_{\ \mu\nu} - \bar{\nabla}_\nu\delta\Gamma^\lambda_{\ \mu\lambda} \right) = \\
    =\int \! \mathrm{d^4}x\ \! \sqrt{-g}\left\{ -\bar{\nabla}_\lambda(\sqrt{-g}f'(\mathcal{R})g^{\mu\nu}) + \bar{\nabla}_\sigma(\sqrt{-g}f'(\mathcal{R})g^{\mu\sigma})\delta^{\,\nu}_{\ \lambda} \right\}\delta\Gamma^\lambda_{\ \mu\nu}
    \end{gathered}
    \end{equation}
    Now we can write the field equation for the Palatini formalism as
    \begin{equation}\label{3.18}
    f'(\mathcal{R})\mathcal{R}_{(\mu\nu)} - \frac{1}{2}f(\mathcal{R})g_{\mu\nu} = \kappa\,T_{\mu\nu}
    \end{equation}
    \begin{equation}\label{3.19}
    -\bar{\nabla}_\lambda(\sqrt{-g}f'(\mathcal{R})g^{\mu\nu}) + \bar{\nabla}_\sigma(\sqrt{-g}f'(\mathcal{R})g^{\sigma(\mu})\delta^{\,\nu)}_{\ \lambda} = 0
    \end{equation}
    From the trace of eq. (\ref{3.19}) we obtain the following relation
    \begin{equation}\label{3.20}
    \bar{\nabla}_\sigma(\sqrt{-g}f'(\mathcal{R})g^{\sigma\mu}) = 0
    \end{equation}
    with the help of which we can rewrite the field equation in a more simple form
    \begin{equation}\label{3.21}
    f'(\mathcal{R})\mathcal{R}_{(\mu\nu)} - \frac{1}{2}f(\mathcal{R})g_{\mu\nu} = \kappa\,T_{\mu\nu}
    \end{equation}
    \begin{equation}\label{3.22}
    \bar{\nabla}_\lambda(\sqrt{-g}f'(\mathcal{R})g^{\mu\nu}) = 0
    \end{equation}
    Standard GR can be obtained in the Palatini formalism if we choose $ f(\mathcal{R}) = \mathcal{R} $, case in which $f'(\mathcal{R}) = 1 $ which will transform relation (\ref{3.22}) into the Levi-Civita connection. If this is the case, then eq. (\ref{3.21}) will become the Einstein equations because now $ \mathcal{R}_{\mu\nu} = R_{\mu\nu}$ and $ \mathcal{R} = R $. Thus we see that in Palatini formalism we obtain the Levi-Civita connection as a dynamical consequences of the theory insted of considering it an a priori assumption.
    
    From the trace of eq. (\ref{3.21}) we obtain
    \begin{equation}\label{3.23}
    f'(\mathcal{R})\mathcal{R} - 2f(\mathcal{R}) = \kappa\,T
    \end{equation}
    which is an algebraic equation in $ \mathcal{R} $. If $ T = 0 $, (vacuum and electrovacuum for example), equation (\ref{3.23}) reduces to
    \begin{equation}\label{3.24}
    f'(\mathcal{R})\mathcal{R} - 2f(\mathcal{R}) = 0
    \end{equation}
    One of the solutions of this equation is $\mathcal{R}=const$. Another interesting solution is the case when $ f(\mathcal{R}) \varpropto \mathcal{R}^2 $ because this will lead to a conformally invariant theory \cite{50}.

    In principle we can eliminate the independent connection between eq. (\ref{3.21} - \ref{3.22}) if we know a solution to eq. (\ref{3.23}). In order to do this, let us first start by making a conformal transformation of the metric $ g $ (for more details on conformal transformations see section 6 )
    \begin{equation}\label{3.25}
    h_{\mu\nu} = f'(\mathcal{R})g_{\mu\nu}
    \end{equation}
    with the help of which one can deduce the following relation 
    \begin{equation}\label{3.26}
    \sqrt{g}f'(\mathcal{R})g^{\mu\nu} = \sqrt{g}h^{\mu\nu}
    \end{equation}
    If we introduce the above relation in eq. (\ref{3.22}) we obtain the definition of the Levi-Civita connection of $ h_{\mu\nu} $. Then one can algebraically solve this new relation to express the independent connection as
    \begin{equation}\label{3.27}
    \Gamma^\lambda_{\ \mu\nu} = h^{\lambda\sigma}\left( \partial_\mu h_{\nu\sigma} + \partial_\nu h_{\mu\sigma} - \partial_\sigma h_{\mu\nu} \right)
    \end{equation}
    or in terms of $ g_{\mu\nu} $ by
    \begin{equation}\label{3.28}
    \Gamma^\lambda_{\ \mu\nu} = \frac{1}{f'(\mathcal{R})}g^{\lambda\sigma}\left\{ \partial_\mu(f'(\mathcal{R})g_{\nu\sigma}) + \partial_\nu (f'(\mathcal{R})g_{\mu\sigma}) - \partial_\sigma (f'(\mathcal{R})g_{\mu\nu}) \right\}
    \end{equation}
    Under a conformal transformation the Ricci tensor Ricci scalar become (see section 6 and \cite{27}, \cite{28})
    \begin{equation}\label{3.29}
    \begin{split}
    \mathcal{R}_{\mu\nu} &= R_{\mu\nu} + \frac{3}{2}\frac{1}{f'(\mathcal{R})^2}\nabla_\mu f'(\mathcal{R})\nabla_\nu f'(\mathcal{R}) - \\
    &-\frac{1}{f'(\mathcal{R})}\left( \nabla_\mu\nabla_\nu - \frac{1}{2}g_{\mu\nu}\Box \right)f'(\mathcal{R})
    \end{split}
    \end{equation}
    \begin{equation}\label{3.30}
    \mathcal{R} = R + \frac{3}{2f'(\mathcal{R})^2}\nabla_\mu f'(\mathcal{R})\nabla^\mu f'(\mathcal{R}) + \frac{3}{f'(\mathcal{R})}\Box f'(\mathcal{R})
    \end{equation}
    Introducing eq. (\ref{3.29}) and (\ref{3.30}) in eq. (\ref{3.21}) we obtain in the end
    \begin{equation}\label{3.31}
    \begin{gathered}
    G_{\mu\nu} = \frac{\kappa}{f'}T_{\mu\nu} - \frac{1}{2}g_{\mu\nu}\left( \mathcal{R} - \frac{f}{f'} \right) + \frac{1}{f'}(\nabla_\mu\nabla_\nu - g_{\mu\nu}\Box)f' - \\
    -\frac{3}{2}\frac{1}{f'^2}\left[ \nabla_\mu f'\nabla_\nu f' - \frac{1}{2}g_{\mu\nu}(\nabla f')^2 \right]
    \end{gathered}
    \end{equation}
    We can regard eq. (\ref{3.31}) as an Einstein equation with modified source. From it we can deduce the following observation

    a) If $ f(\mathcal{R}) = \mathcal{R} $, then the theory will transform into GR.

    b) For matter fields with $ T = 0 $ the theory becomes GR with a cosmological constant given by
    \begin{equation}\label{3.32}
    \Lambda = \frac{\mathcal{R}_0}{4}
    \end{equation}
    where $\mathcal{R}_0$ is the value of $\mathcal{R}$ in the case of $T=0$ and we also used eq. (\ref{3.24}) in the derivation of $\Lambda$. 

    c) In the cases for which $ T \neq 0 $, the RHS of eq. (\ref{3.31}) will include also derivatives of the stress-energy tensor, which are absent in GR.

\section{Boundary term in metric f(R) theories}\label{1.4}

    The action for $f(R)$ gravity (without the matter term) reads
    \begin{equation}\label{b1}
    \mathcal{S} = \int\!\mathrm{d^4}x\,\sqrt{g}f(R)
    \end{equation}
    In order to obtain the field equation corresponding to this action one must compute the variation of action (\ref{b1}) with respect to the metric. This will lead us to the following result
    \begin{equation}\label{b2}
    \begin{split}
    \delta\mathcal{S} = \int\!\mathrm{d^4}x\,\sqrt{g}\, &\left\{ f'(R)R_{\mu\nu} -\frac{1}{2}f(R)g_{\mu\nu}-
    \nabla_\mu\nabla_\nu f'(R)+g_{\mu\nu}\Box f'(R) \right\}\,\delta g^{\mu\nu}\\
    &-\oint_{\partial\Sigma}\!\mathrm{d^{3}}y\,\sqrt{|h|}\,\epsilon\,f'(R) n^\lambda h^{\mu\nu}\partial_\lambda(\delta g_{\mu\nu})
    \end{split}
    \end{equation}
    Like in the case of Einstein-Hilbert action the last term in (\ref{b2}) represents a boundary surface term. Let us denote it by
    \begin{equation}\label{b3}
    \delta\mathcal{S'}_{B} =-\oint_{\partial\Sigma}\!\mathrm{d^{3}}y\,\sqrt{|h|}\,\epsilon\,f'(R) n^\lambda h^{\mu\nu}\partial_\lambda(\delta g_{\mu\nu})
    \end{equation}
    It is important to stress out that this surface term can not be written as the total variation of a quantity, due to the presence of $f'(R)$. Because of this fact the action (\ref{b1}) can not be fixed (like in the case of Einstein-Hilbert action) by subtracting a suitable surface term before making the variation. This type of boundary term appears in the majority of higher-order theories of gravity (see for example \cite{38}, \cite{46}, \cite{47}, \cite{31}, \cite{48},\cite{51}). A brief discussion about boundary terms for general scalar-tensor theories can be found in \cite{49}.

    The surface boundary term (\ref{b3}) can also be written as a Gibbons-York-Hawking surface term of the form
    \begin{equation}\label{b4}
    \mathcal{S'}_{GYH}=2\oint_{\partial\Sigma}\!\mathrm{d^{3}}y\,\sqrt{|h|}\,\epsilon f'(R) K
    \end{equation}
    Using (\ref{49}) for the variation of the extrinsic curvature $K$ and
    \begin{equation}\label{b5}
    \delta f(R)=f'(R)\delta R
    \end{equation}
    the boundary term (\ref{b4}) transforms as
    \begin{equation}\label{b6}
    \begin{split}
    \delta\mathcal{S'}_{GYH}&=2\oint_{\partial\Sigma}\!\mathrm{d^{3}}y\,\sqrt{|h|}\,\epsilon\left\{ K\delta f'(R) + f'(R)\delta K \right\}\\
    &=2\oint_{\partial\Sigma}\!\mathrm{d^{3}}y\,\sqrt{|h|}\,\epsilon\left\{ f''(R)K\delta R + f'(R)\delta K \right\}\\
    &=2\oint_{\partial\Sigma}\!\mathrm{d^{3}}y\,\sqrt{|h|}\,\epsilon f''(R)K\delta R + \oint_{\partial\Sigma}\!\mathrm{d^{3}}y\,\sqrt{|h|}\,\epsilon\,f'(R) n^\lambda h^{\mu\nu}\partial_\lambda(\delta g_{\mu\nu})
    \end{split}
    \end{equation}
    From this we see that indeed the presence of a boundary surface term in the initial action it is not enough to make it stationary. For that to happen we must also impose $\delta R=0$ on the boundary \cite{31}.

    In what follows we will present a derivation of eq. (\ref{b2}). The variation of action (\ref{b1}) is given by
    \begin{equation}\label{b7}
    \delta\mathcal{S} = \int\!\mathrm{d^4}x\,\left\{ \delta(\sqrt{g})f(R)+\sqrt{g}\delta f(R) \right\}
    \end{equation}
    Using eq. (\ref{b5}),(\ref{15}),(\ref{23}) and taking into account that
    \begin{equation}\label{b8}
    \delta R=\delta(g^{\mu\nu}R_{\mu\nu})=\delta(g^{\mu\nu})R_{\mu\nu}+g^{\mu\nu}\delta R_{\mu\nu}
    \end{equation}
    the variation of action (\ref{b7}) transforms in
    \begin{equation}\label{b9}
    \delta\mathcal{S} = \int\!\mathrm{d^4}x\,\sqrt{g}\left\{ f'(R)R_{\mu\nu}\delta g^{\mu\nu} - \frac{1}{2}g_{\mu\nu}f(R)\delta g^{\mu\nu} +f'(R)\nabla_\nu(g^{\mu\nu}\delta\Gamma^\sigma_{\ \mu\sigma}-g^{\mu\sigma}\delta\Gamma^\nu_{\ \mu\sigma}) \right\}
    \end{equation}
    Next let us further transform the last term in integral (\ref{b9}). The first step is to calculate the variation of $\Gamma$, which, starting from the definition, will be
    \begin{equation}\label{b10}
    \delta\Gamma^\sigma_{\ \mu\nu} = \frac{1}{2}\delta g^{\sigma\lambda}\left( \partial_\mu g_{\lambda\nu}+\partial_\nu g_{\mu\lambda}-\partial_\lambda g_{\mu\nu} \right) + \frac{1}{2}g^{\sigma\lambda}\left\{ \partial_\mu(\delta g_{\lambda\nu})+\partial_\nu (\delta g_{\mu\lambda})-\partial_\lambda (\delta g_{\mu\nu}) \right\}
    \end{equation}
    It is always possible to work in a local coordinate system in which $\Gamma=0$ at a given fixed point $P$ and at this point the ordinary derivative is identical with the covariant derivative ($\partial\equiv\nabla$). This choice will not affect the final result for $\delta\Gamma$ because this quantity is a tensor and in the end will have the same expression in any coordinate system. Taking this into account eq. (\ref{b10}) becomes
    \begin{equation}\label{b11}
    \begin{split}
    \delta\Gamma^\sigma_{\ \mu\nu} &= \frac{1}{2}\delta g^{\sigma\lambda}\left( \nabla_\mu g_{\lambda\nu}+\nabla_\nu g_{\mu\lambda}-\nabla_\lambda g_{\mu\nu} \right) + \frac{1}{2}g^{\sigma\lambda}\left\{ \nabla_\mu(\delta g_{\lambda\nu})+\nabla_\nu (\delta g_{\mu\lambda})-\nabla_\lambda (\delta g_{\mu\nu}) \right\}\\
    &=\frac{1}{2}g^{\sigma\lambda}\left\{ \nabla_\mu(\delta g_{\lambda\nu})+\nabla_\nu (\delta g_{\mu\lambda})-\nabla_\lambda (\delta g_{\mu\nu}) \right\}
    \end{split}
    \end{equation}
    where in the last line we used the metricity of the metric $\nabla_\mu g_{\lambda\nu}=0$.

    We can now compute the quantity
    \begin{equation}\label{b12}
    \begin{split}
    g^{\mu\nu}\delta\Gamma^\sigma_{\ \mu\sigma}&=\frac{1}{2}\nabla_\mu(g^{\mu\nu}g^{\sigma\lambda}\delta g_{\lambda\sigma})+ \frac{1}{2}\nabla_\sigma(g^{\mu\nu}g^{\sigma\lambda}\delta g_{\lambda\mu}) -\frac{1}{2}\nabla_\lambda(g^{\mu\nu}g^{\sigma\lambda}\delta g_{\mu\sigma})\\
    &=\frac{1}{2}\nabla_\mu(g^{\mu\nu}g^{\sigma\lambda}\delta g_{\lambda\sigma}) - \frac{1}{2}\nabla_\sigma(\delta g^{\nu\sigma})+\frac{1}{2}\nabla_\lambda(\delta g^{\nu\lambda})\\
    &=-\frac{1}{2}\nabla_\mu(g_{\mu\nu}g^{\mu\nu}\delta g^{\mu\nu})
    \end{split}
    \end{equation}
    In the second and last lines of eq. (\ref{b12}) we've relabeled some indices and also we used the identities
    \begin{equation}\label{b13}
    \delta g^{\nu\sigma}=-g^{\mu\nu}g^{\sigma\lambda}\delta g_{\mu\lambda} \ \ \ \ \ \ g^{\mu\nu}\delta g_{\\mu\nu}=-g_{\mu\nu}\delta g^{\mu\nu}
    \end{equation}
    Following the same steps we also obtain
    \begin{equation}\label{b14}
    g^{\mu\sigma}\delta\Gamma^\nu_{\ \mu\sigma}=-\nabla_\mu\delta g^{\mu\nu}+ \frac{1}{2}\nabla_\mu(g_{\mu\nu}\delta g^{\mu\nu}g^{\mu\nu})
    \end{equation}
    In the end we have
    \begin{equation}\label{b15}
    \nabla_\nu(g^{\mu\nu}\delta\Gamma^\sigma_{\ \mu\sigma}-g^{\mu\sigma}\delta\Gamma^\nu_{\ \mu\sigma})=\nabla_\mu(\delta g^{\mu\nu}) - g_{\mu\nu}g^{\mu\nu}\nabla_\mu(\delta g^{\mu\nu})
    \end{equation}
    Introducing the above result in eq. (\ref{b9}) will give for the variation of the action the next expression
    \begin{equation}\label{b16}
    \begin{split}
    \delta\mathcal{S} &= \int\!\mathrm{d^4}x\,\sqrt{g}\left( f'(R)R_{\mu\nu} - \frac{1}{2}g_{\mu\nu}f(R)\right)\delta g^{\mu\nu} \\
    &+ \int\!\mathrm{d^4}x\,\sqrt{g}\bigg( f'(R)\nabla_\mu\nabla_\nu(\delta g^{\mu\nu}) - f'(R)g_{\mu\nu}\Box(\delta g^{\mu\nu})\bigg)
    \end{split}
    \end{equation}
    We observe that the two terms from the last integral of (\ref{b16}) can be written in the following way
    \begin{equation}\label{b17}
    f'(R)\nabla_\mu\nabla_\nu(\delta g^{\mu\nu})=\nabla_\mu\nabla_\nu(f'(R)\delta g^{\mu\nu})- \delta g^{\mu\nu}\nabla_\mu\nabla_\nu(f'(R))
    \end{equation}
    \begin{equation}\label{b18}
    f'(R)g_{\mu\nu}\Box(\delta g^{\mu\nu})=\Box(f'(R)g_{\mu\nu}\delta g^{\mu\nu})-\delta g^{\mu\nu}g_{\mu\nu}\Box(f'(R))
    \end{equation}
    We will use this results to calculate the next integrals
    \begin{equation}\label{b19}
    I=\int\!\mathrm{d^4}x\,\sqrt{g} f'(R)\nabla_\mu\nabla_\nu(\delta g^{\mu\nu})=\int\!\mathrm{d^4}x\,\sqrt{g}\,\nabla_\mu\nabla_\nu(f'(R)\delta g^{\mu\nu} - \int\!\mathrm{d^4}x\,\sqrt{g}\,\delta g^{\mu\nu}\nabla_\mu\nabla_\nu f'(R)
    \end{equation}
    Using the Gauss-Stokes theorem (\ref{39}) on the first integral of eq. (\ref{b19}), I becomes
    \begin{equation}\label{b20}
    I=\oint_{\partial\Sigma}\!\mathrm{d^3}y\,\sqrt{|h|}\,\epsilon\,n_\mu A^\mu-\int\!\mathrm{d^4}x\,\sqrt{g}\,\delta g^{\mu\nu}\nabla_\mu\nabla_\nu f'(R)
    \end{equation}
    where we've made the notation
    \begin{equation}\label{b21}
    A^\mu=\nabla_\nu(f'(R)\delta g^{\mu\nu})
    \end{equation}
    In an analogous way
    \begin{equation}\label{b22}
    J=\oint_{\partial\Sigma}\!\mathrm{d^3}y\,\sqrt{|h|}\,\epsilon\,n^\mu B_\mu-\int\!\mathrm{d^4}x\,\sqrt{g}\,\delta g^{\mu\nu}g_{\mu\nu}\Box f'(R)
    \end{equation}
    where
    \begin{equation}\label{b23}
    B_\mu=\nabla_\mu(f'(R)g_{\mu\nu}\delta g^{\mu\nu})
    \end{equation}
    One of the last steps in obtaining a final expression for the variation of action (\ref{b1}) is to calculate the values of the quantities $ n_\mu A^\mu$ and $n^\mu B_\mu$ on the boundary $\partial\Sigma$ .Taking into account that on the boundary $\delta g^{\mu\nu}=0$ we then have
    \begin{equation}\label{b24}
    \begin{split}
    n_\mu A^\mu\bigg|_{\partial\,\Sigma}&=n_\mu\nabla_\nu(f'(R)\delta g^{\mu\nu})n_\mu\nabla_\nu(f'(R))\delta g^{\mu\nu} +n_\mu f'(R)\nabla_\nu(\delta g^{\mu\nu})\\
    &=n_\mu f'(R)\nabla_\nu(\delta g^{\mu\nu})\\
    &=-n_\mu f'(R) g^{\mu\sigma}g^{\nu\lambda}\nabla_\nu(g_{\sigma\lambda})\\
    &=-n_\mu f'(R)(h^{\mu\sigma}+\epsilon n^\mu n^\sigma)(h^{\nu\lambda}+\epsilon n^\nu n^\lambda)\nabla_\nu(g_{\sigma\lambda})\\
    &=-f'(R)(n_\mu h^{\mu\sigma}+\epsilon n_\mu n^\mu n^\sigma)(h^{\nu\lambda}+\epsilon n^\nu n^\lambda)\nabla_\nu(g_{\sigma\lambda})\\
    &=-f'(R)(\epsilon n^\sigma h^{\nu\lambda}\nabla_\nu(\delta g_{\sigma\lambda})-\epsilon^2n^\sigma n^\nu n^\lambda\nabla_\nu(\delta g_{\sigma\lambda}))=0
    \end{split}
    \end{equation}
    where we used the proprieties $n_\sigma h^{\sigma\mu}=0$, $\epsilon^2=1$ and the fact that on the boundary the tangential derivative of the metric vanishes $ h^{\nu\lambda}\nabla_\nu(\delta g_{\sigma\lambda})$.
    \begin{equation}\label{b25}
    \begin{split}
    n^\mu B_\mu\bigg|_{\partial\,\Sigma}&=n^\mu\nabla_\mu(f'(R)g_{\mu\nu}\delta g^{\mu\nu})\\
    &= n^\mu\nabla_\mu(f'(R))g_{\mu\nu}\delta g^{\mu\nu} +n^\mu\nabla_\mu(g_{\mu\nu})f'(R)\delta g^{\mu\nu}+ n^\mu f'(R)g_{\mu\nu}\nabla_\mu(\delta g^{\mu\nu})\\
    &=n^\mu f'(R)g_{\mu\nu}\nabla_\mu(\delta g^{\mu\nu}\\
    &=-n^\mu f'(R)g^{\mu\nu}\nabla_\mu(\delta g_{\mu\nu}\\
    &=-n^\mu f'(R)(h^{\mu\nu}+\epsilon n^\mu n^\nu)\nabla_\mu(\delta g_{\mu\nu}\\
    &=-f'(R)n^\mu h^{\mu\nu}\nabla_\mu(\delta g_{\mu\nu}
    \end{split}
    \end{equation}
    Putting all together we arrive at the final result given by eq. (\ref{b2}).

\section{Conformal transformations}

    The field equations derived in the metric formalism of $ f(R) $ theory can be put into a simplified form by performing a conformal transformation on the metric. In this conformally transformed frame the field equations will have the same form as the Einstein equation, with a minimally coupled scalar field. For this reason, the new frame is called the Einstein frame, while the original frame is known as the Jordan frame. There is a debate on which frame is the physical one, The Einstein or the Jordan frame? Most of the scientists tend to consider the original Jordan frame to be the physical one. However, there are reasons to believe that the Einstein frame is the one with a physical meaning. A detailed discussion regarding this problem can be found in \cite{14},\cite{15}. For conformal transformations used in $f(R)$ theories, and scalar-tensor theories in general, the reader can consult \cite{14}, \cite{16}, \cite{17}, \cite{18}, \cite{21}, \cite{22}, \cite{25}, \cite{26} among others.

    For a general spacetime $ (M,g_{\mu\nu}) $ the following transformation performed on the metric
    \begin{equation}\label{c1}
    \tilde g_{\mu\nu}=\Omega^2g_{\mu\nu}
    \end{equation}
    where $\Omega$ ia a conformal factor (a nonvanishing, regular function) is called a Weil or conformal transformation. The quantities with a tilde will represent quantities in the Einstein frame. It can be shown that the spacetimes $ (M,g_{\mu\nu})$ and $(M,\tilde g_{\mu\nu})$ have the same causal structure \cite{3}.

    The action for $f(R)$ gravity in the metric formalism is
    \begin{equation}\label{c2}
    \mathcal{S}_{met} = \frac{1}{2\kappa^2}\int\!\mathrm{d^4}x\,\sqrt{g}f(R) + \int\!\mathrm{d^4}x\,\mathcal{L}_M(g_{\mu\nu},\psi)
    \end{equation}
    If we add and subtract $ R F(R) $ in the first term of the action (\ref{c2}), then we can rewrite the action as
    \begin{equation}\label{c3}
    \mathcal{S}_{met} = \int\!\mathrm{d^4}x\,\sqrt{g}\left( \frac{1}{2\kappa^2}R F(R)-Y \right) + \int\!\mathrm{d^4}x\,\mathcal{L}_M(g_{\mu\nu},\psi)
    \end{equation}
    where we denoted by $ Y $ the following quantity
    \begin{equation}\label{c4}
    Y=\frac{1}{2\kappa^2}(R F(R)-f)
    \end{equation}
    The relations for the Ricci scalars and for the determinants in the two frames are \cite{27}, \cite{28}
    \begin{equation}\label{c5}
    \sqrt{g}=\Omega^{-4}\sqrt{\tilde g}
    \end{equation}
    \begin{equation}\label{c6}
    R=\Omega^2[ \tilde R + 6\Box(\ln\Omega)-6\tilde g^{\mu\nu}\nabla_\mu(\ln\Omega)\nabla_\nu(\ln\Omega)  ]
    \end{equation}
    Using these relations, the action (\ref{c3}) is transformed as
    \begin{equation}\label{c7}
    \begin{split}
    \mathcal{S}_{met} &= \int\!\mathrm{d^4}x\,\Omega^{-4}\sqrt{\tilde g}\left\{ \frac{1}{2\kappa^2}F(R)\Omega^2[ \tilde R + 6\Box(\ln\Omega)-6\tilde g^{\mu\nu}\nabla_\mu(\ln\Omega)\nabla_\nu(\ln\Omega)  ]-Y \right\} \\ &+\int\!\mathrm{d^4}x\,\mathcal{L}_M(\Omega^{-2}\tilde g_{\mu\nu},\psi)\\
    &=\int\!\mathrm{d^4}x\,\sqrt{\tilde g}\left\{ \frac{1}{2\kappa^2}F(R)\Omega^{-2}[ \tilde R + 6\Box(\ln\Omega)-6\tilde g^{\mu\nu}\nabla_\mu(\ln\Omega)\nabla_\nu(\ln\Omega)  ]-\Omega^{-4}Y \right\}\\  &+\int\!\mathrm{d^4}x\,\mathcal{L}_M(\Omega^{-2}\tilde g_{\mu\nu},\psi)
    \end{split}
    \end{equation}
    If we want the action (\ref{c7}) to be linear in $ \tilde R $, then we must impose the condition
    \begin{equation}\label{c8}
    F(R)=\Omega^2
    \end{equation}
    thus obtaining the Einstein frame action
    \begin{equation}\label{c9}
    \begin{split}
    \mathcal{S}_{met} &= \int\!\mathrm{d^4}x\,\sqrt{\tilde g}\left\{ \frac{1}{2\kappa^2}[ \tilde R + 6\Box(\ln\Omega)-6\tilde g^{\mu\nu}\nabla_\mu(\ln\Omega)\nabla_\nu(\ln\Omega)  ]-\frac{Y}{F(R)^2} \right\} \\ &+\int\!\mathrm{d^4}x\,\mathcal{L}_M(F^{-1}\tilde g_{\mu\nu},\psi)
    \end{split}
    \end{equation}
    Using the Gauss theorem it is straightforward to show that
    \begin{equation}\label{c10}
    \int\!\mathrm{d^4}x\,\sqrt{\tilde g}\frac{1}{2\kappa^2}6\Box(\ln\Omega)=0
    \end{equation}
    Taking this into account the action (\ref{c9}) becomes
    \begin{equation}\label{c11}
    \begin{split}
    \mathcal{S}_{met} &= \int\!\mathrm{d^4}x\,\sqrt{\tilde g}\left\{ \frac{1}{2\kappa^2}[ \tilde R -6\tilde g^{\mu\nu}\nabla_\mu(\ln\Omega)\nabla_\nu(\ln\Omega)  ]-\frac{Y}{F(R)^2} \right\} \\ &+\int\!\mathrm{d^4}x\,\mathcal{L}_M(F^{-1}\tilde g_{\mu\nu},\psi)
    \end{split}
    \end{equation}
    We introduce a new scalar field $\phi$ defined as
    \begin{equation}\label{c12}
    \phi\equiv\frac{1}{\kappa}\sqrt{\frac{3}{2}}\ln F(R)
    \end{equation}
    from which follows that
    \begin{equation}\label{c13}
    F(R)=\exp\left( \kappa\sqrt{\frac{2}{3}}\phi \right)
    \end{equation}
    From $F(R)=\Omega^2$ after logaritmation we have $\ln F(R)=2\ln\Omega$. Introducing the new scalar field into the action we obtain
    \begin{equation}\label{c14}
    \mathcal{S}_{met} = \int\!\mathrm{d^4}x\,\sqrt{\tilde g}\left\{ \frac{1}{2\kappa^2}\tilde R -\frac{1}{2}\tilde g^{\mu\nu}\nabla_\mu\phi\nabla_\nu\phi  -V(\phi) \right\}+\int\!\mathrm{d^4}x\,\mathcal{L}_M(F^{-1}(\phi)\tilde g_{\mu\nu},\psi)
    \end{equation}
    where $V(\phi)$ represents the potential of the scalar field and it is defined as
    \begin{equation}\label{c15}
    V(\phi)=\frac{Y}{F(R)^2}=\frac{R F(R)-f}{2\kappa^2F(R)^2}
    \end{equation}
    From the second term of action (\ref{c14}) we can conclude that the scalar field $\phi$ is directly coupled to matter in the Einstein frame. This fact is more visible after calculating the variation of action (\ref{c14}) with respect to the scalar field $\phi$
    \begin{equation}\label{c16}
    -\partial_\mu\left( \frac{\partial(\sqrt{\tilde g}\mathcal{L}_{\phi})}{\partial(\partial_\mu\phi)} \right) + \frac{\partial(\sqrt{\tilde g}\mathcal{L}_{\phi})}{\partial\phi} + \frac{\partial{\mathcal{L}_{M}}}{\partial\phi}=0
    \end{equation}
    where $ \mathcal{L}_{\phi} $ represents the lagrangean density of the scalar filed $ \phi $ and is given by
    \begin{equation}\label{c17}
    \mathcal{L}_{\phi}=-\frac{1}{2}\tilde g^{\mu\nu}\nabla_\mu\phi\nabla_\nu\phi  -V(\phi)
    \end{equation}
    from which we can calculate it's energy-momentum tensor
    \begin{equation}\label{c18}
    \tilde T^{(\phi)}_{\mu\nu}=\frac{-2}{\sqrt{\tilde g}}\frac{\delta(\sqrt{\tilde g}\mathcal{L}_{\phi})}{\delta \tilde g^{\mu\nu}}= \nabla_\mu\phi\nabla_\nu\phi-\tilde g_{\mu\nu}\left\{ \frac{1}{2}\tilde g^{\alpha\beta}\nabla_\alpha\phi\nabla_\beta\phi +V(\phi) \right\}
    \end{equation}
    Now we will compute the first two terms of the equation (\ref{c16}
    \begin{equation}\label{c19}
    -\partial_\mu\left( \frac{\partial(\sqrt{\tilde g}\mathcal{L}_{\phi})}{\partial(\partial_\mu\phi)} \right)=\partial_\mu(\sqrt{\tilde g}\,\tilde g^{\mu\nu}\nabla_\nu\phi)
    \end{equation}
    \begin{equation}\label{c20}
    \frac{\sqrt{\tilde g}\mathcal{L}_{\phi})}{\partial\phi}=-\sqrt{\tilde g}\,\frac{\partial V(\phi)}{\partial\phi}
    \end{equation}
    Introducing this back into (\ref{c16}) we get
    \begin{equation}\label{c21}
    \partial_\mu(\sqrt{\tilde g}\,\tilde g^{\mu\nu}\nabla_\nu\phi)-\sqrt{\tilde g}\,\frac{\partial V(\phi)}{\partial\phi} +\frac{\partial{\mathcal{L}_{M}}}{\partial\phi}=0
    \end{equation}
    that is equivalent to
    \begin{equation}\label{c22}
    \tilde\Box\phi-\frac{\partial V(\phi)}{\partial\phi}+ \frac{1}{\sqrt{\tilde g}}\frac{\partial{\mathcal{L}_{M}}}{\partial\phi}=0
    \end{equation}
    where
    \begin{equation}\label{c23}
    \tilde\Box=\frac{1}{\sqrt{\tilde g}}\partial_\mu(\sqrt{\tilde g}\,\tilde g^{\mu\nu}\nabla_\nu)
    \end{equation}
    The last term of eq. (\ref{c22}) can be further transformed as
    \begin{equation}\label{c24}
    \begin{split}
    \frac{\partial{\mathcal{L}_{M}}}{\partial\phi}&= \frac{\delta{\mathcal{L}_{M}}}{\delta g^{\mu\nu}}\frac{\partial g^{\mu\nu}}{\partial\phi} =-\frac{\sqrt{g}}{2}\left(\frac{-2}{\sqrt{g}}\frac{\delta{\mathcal{L}_{M}}}{\delta g^{\mu\nu}}\right)\frac{\partial g^{\mu\nu}}{\partial\phi}\\
    &= -\frac{\sqrt{g}}{2}T^{(M)}_{\mu\nu}\frac{\partial g^{\mu\nu}}{\partial\phi}= -\sqrt{\tilde g}\frac{1}{2F(R)}\frac{\partial F(R)}{\partial\phi}\tilde g^{\mu\nu}\tilde T^{(M)}_{\mu\nu}
    \end{split}
    \end{equation}
    In the last line of eq. (\ref{c24}) we used the relations $ \sqrt{\tilde g}=F^2(R)\sqrt{g} $ and $ g^{\mu\nu}=F(R)\tilde g^{\mu\nu}$.
    Also the energy-momentum tensor of the matter field can be transformed in the following way
    \begin{equation}\label{c25}
    \tilde T^{(M)}_{\mu\nu}=\frac{-2}{\sqrt{\tilde g}}\frac{\delta(\mathcal{L}_{M})}{\delta \tilde g^{\mu\nu}}= \frac{-2}{F(R)^2\sqrt{\tilde g}}\frac{\delta(\mathcal{L}_{M})}{F^{-1}(R)\delta \tilde g^{\mu\nu}}=\frac{T^{(M)}_{\mu\nu}}{F(R)}
    \end{equation}
    We define the strength of the coupling between the field and matter by
    \begin{equation}\label{c26}
    Q=\frac{-1}{2\kappa F(R)}\frac{\partial F(R)}{\partial\phi}
    \end{equation}
    It then follows that
    \begin{equation}\label{c27}
    \frac{\partial{\mathcal{L}_{M}}}{\partial\phi}=\sqrt{\tilde g}\,\kappa Q\tilde T^{(M)}
    \end{equation}
    where $ \tilde T $ stands for the trace of the matter energy-momentum tensor.

    Introducing this into eq. (\ref{c22}), the Klein-Gordon type equation in the Einstein frame will be
    \begin{equation}\label{c28}
    \tilde\Box\phi-\frac{\partial V(\phi)}{\partial\phi}+\kappa Q\tilde T=0
    \end{equation}
    which shows explicitly the direct coupling between the matter and the scalar field $ \phi $ (apart from radiation when $\tilde T=0$).

\section{Friedman equations and classical cosmology}

    \subsection{The cosmological Principle}

    The most of the luminous matter contained in the Universe can be found in stars that form more complex structures like galaxies \cite{66}. Astronomical observations showed that galaxies form much bigger structures known under the name of clusters of galaxies. There are hints about the existence of mega-clusters of galaxies, but the existence of mega-mega-clusters and so on seems very improbable \cite{20}, \cite{33}, \cite{58}. Although on small scales (the Solar System, nearby galaxies) the distribution of matter is very nonuniform, it is believed that at the scale of the entire Universe the distribution of matter is very uniform i.e. the Universe is homogenous. There also exist strong evidence (like the CMB radiation discovered by Penzias and Wilson \cite{59}) for the isotropy of space, which says that the Universe looks the same in every direction we look. This two proprieties of homogeneity and isotropy form the Cosmological Principle which says: that at any particular time the universe looks the same from all positions in space and all directions in space at any point are equivalent. All the cosmological observations made so far tell us that the Universe we live in is one in which the Cosmological Principle holds. However, there were and are models in which the Cosmological Principle is violated, see for example the hierarchical model proposed by Charlier \cite{60}, the steady state model proposed by Bondy and Gold \cite{61} and at the same time independently by Hoyle \cite{62} (all this models assume the validity of a "perfect cosmological principle" which says that at any time the Universe is homogeneous and isotropic); there is also a proposal made by Pieronero \cite{63} for a fractal type Universe.

    The proprieties of homogeneity and isotropy can be formulated in a much more precise mathematical way \cite{64}, \cite{3}, \cite{8}. Thus we can say that a space M (a differential manifold) is isotropic in the vicinity of a point P if for any two given vectors $v$ and $w$, from the tangent space, there exists an isometry transformation on M such that if we transport the vector $w$ along the isometry it will remain parallel with $v$. The property of homogeneity is related to the fact that the metric defined on the space M has the same form in every point. In other words, for every two points P and Q in M there exists an isometry that can move point P into the point Q.

    In general, it is not necessarily to exist a connection between the proprieties of homogeneity and isotropy of a given manifold M. For example, a manifold can be homogeneous in every point but not isotropic ($R\times S^2$ with the standard metric) or it can be isotropic around a point and not homogeneous (the cone is an example). If a manifold is isotropic around every point then it is also homogeneous; and also if it is isotropic around a point and homogeneous in the same time then it will be isotropic around every point.

    \subsection{Einstein Equations for FRW metric}

    It can be shown that for a Universe (manifold) in which the Cosmological Principle is valid the metric of this manifold it's always the Friedman-Robertson-Walker (FRW) metric \cite{1}, \cite{8}, \cite{65}
    \begin{equation}\label{lic1}
    ds^2=-dt^2+a^2(t)\left[\frac{dr^2}{1-kr^2}+r^2(d\theta^2+\sin^2\theta d\phi^2)\right]
    \end{equation}
    where $a(t)$ is the scale factor and $k$ give us the spatial curvature: negative ($k=-1$), flat ($k=0$) or positive ($k=+1$).
    We will assume that the matter distribution in the Universe is of a perfect cosmological fluid form, with the following energy-momentum tensor \cite{8}
    \begin{equation}\label{lic2}
    T^{\mu\nu}=\left(\rho+p\right)u^\mu u^\nu+pg^{\mu\nu}
    \end{equation}
    where $\rho$ represents the density of the cosmological fluid and $p$ stands for it's pressure and $u^{\mu}$ is the velocity of the fluid in comoving frame given by
    \begin{equation}\label{lic3}
    u^{\mu}=(1,0,0,0)
    \end{equation}
    Due to the fact that our space is homogenous and isotropic the density and pressure of the cosmological fluid can be functions only on time, being independent of the spatial coordinates.

    In order to write the Einstein field equations for gravity we must first calculate the Ricci tensor and Ricci scalar curvature with the help of the following formulae
    \begin{equation}\label{lic4}
    \begin{split}
    &R_{\mu\nu}=\Gamma^\lambda_{\ \mu\nu,\lambda}-\Gamma^\lambda_{\ \mu\lambda,\nu}+\Gamma^\lambda_{\ \mu\nu}\Gamma^\sigma_{\ \lambda\sigma}-\Gamma^\sigma_{\ \mu\lambda}\Gamma^\lambda_{\ \nu\sigma}\\
    &\Gamma^\sigma_{\ \mu\nu}=\frac{1}{2}g^{\sigma\rho}(g_{\rho\mu,\nu}+g_{\rho\nu,\mu}-g_{\mu\nu,\rho})\\
    &R=g^{\mu\nu}R_{\mu\nu}
    \end{split}
    \end{equation}
    After straight away calculations we obtain the following non-vanishing terms for the Ricci tensor
    \begin{equation}\label{lic5}
    \begin{split}
    &R_{00}=3\frac{\dot a^2+k}{a^2} \\
    &R_{11}=\frac{1}{1-kr^2}(a\ddot a+2\dot a^2+2k) \\
    &R_{22}=r^2(a\ddot a+2\dot a^2+2k) \\
    &R_{33}=r^2\sin^2\theta(a\ddot a+2\dot a^2+2k)
    \end{split}
    \end{equation}
    and the Ricci scalar curvature reads
    \begin{equation}\label{lic6}
    R=\frac{6}{c^2a^2}(a\ddot a+2\dot a^2+2kc^2)
    \end{equation}
    The energy-momentum tensor (\ref{lic2}) in the FRW (\ref{lic1}) anzat will give us the following terms
    \begin{equation}\label{lic7}
    \begin{split}
    &T_{00}=\rho c^4 \ \ \ \ \ \ \ \ T_{11}=\frac{pa^2}{1-kr^2} \\
    &T_{22}=pr^2a^2 \ \ \ \ \ \ \ \ T_{33}=pr^2a^2\sin^2\theta
    \end{split}
    \end{equation}
    Introducing all the above in the Einstein equations (\ref{31}) we obtain for the time-time component the following field equation
    \begin{equation}\label{liv8}
    \dot a^2+k=\frac{8\pi G}{3}\rho a^2
    \end{equation}
    and all the space-space equations are equivalent with the following field equation
    \begin{equation}\label{lic9}
    2a\ddot a+\dot a^2+k=-\frac{8\pi G}{3}pa^2
    \end{equation}
    If we take into account also the cosmological constant $\Lambda$ the new equations will be
    \begin{equation}\label{lic10}
    \begin{split}
    &\ddot a=-\frac{4\pi G}{3}\left(\rho+3p\right)a+\frac{1}{3}\Lambda a \\
    &\dot a^2=\frac{8\pi G}{3}\rho a^2+\frac{1}{3}\Lambda a^2-k
    \end{split}
    \end{equation}

    \subsection{Equation of motion for the cosmological fluid}

    In the previous section we assumed that the matter contained in the Universe can be described with the help of a cosmological fluid which has the energy-momentum tensor given by relation (\ref{lic2}). We can write an equation of motion for the cosmological fluid, which can be obtained in different ways. Here we will use the conservation of the energy-momentum tensor
    \begin{equation}\label{lic11}
    T^{\mu\nu}_{\ \ ;\nu}=0
    \end{equation}
    which can also be written (if we expand the covariant derivative) as
    \begin{equation}\label{lic12}
    T^{\mu\nu}_{\ \ ,\nu}+\Gamma^\mu_{\ \nu\sigma}T^{\sigma\nu}+\Gamma^\nu_{\ \nu\sigma}T^{\mu\sigma}
    \end{equation}
    Next we use the non-vanishing components (\ref{lic7}) of $T^{\mu\nu}$ which will lead us in the end to
    \begin{equation}\label{lic12}
    \dot\rho+\left(\rho+p\right)\frac{3\dot a}{a}=0
    \end{equation}
    for the time component, from which we deduce that the trajectories of the fluid particles (galaxies) are geodesics. As a mater of fact, the eq. (\ref{lic12}) can be also obtained by eliminating $\ddot a$ between eq. (\ref{lic10}) \cite{1}.

    Equation (\ref{lic12}) can be further transformed as
    \begin{equation}\label{lic13}
    \frac{d}{da}(\rho a^3)=-3pa^2 ,
    \end{equation}
    from which we can deduce a relation between the scale factor $a$, density $\rho$ and the pressure $p$. Let us now assume (like in the classical thermodynamics) that each component of the cosmological fluid satisfies the following equation of state
    \begin{equation}\label{lic14}
    p=w\rho
    \end{equation}
    where $w$ is a constant parameter of state (there exists more complex models in which $w$ can depend also on time). The parameter of state $w$ usually takes values between $-1$ and $1$, here are some typical values: $w=0$ for pressureless dust, $w=1/3$ for radiation and $w=-1$ for the cosmological constant.

    Introducing (\ref{lic14}) in eq. (\ref{lic13}) we obtain the following solution
    \begin{equation}\label{lic15}
    \rho=\rho_0\left(\frac{a_0}{a}\right)^{3(1+w)}
    \end{equation}
    where $\rho_0$ and $a_0$ are the values at the time "now" (the moment when the observations are made).

    \subsection{Components of the cosmological fluid}

    In the most general cosmological model the Universe is field, at the same time, with matter and radiation and the cosmological constant is nonzero. Thus we will consider that our cosmological fluid contains three components: matter, radiation and a cosmological constant. We will also assume that the three components do not interact between each others, which is a reasonable assumption that is valid for almost the entire history of the Universe. However, in the early Universe there was an interaction between matter and radiation. Thus, tacking the above into account we can write the total mass equivalent density as
    \begin{equation}\label{lic16}
    \rho(t)=\rho_m(t)+\rho_r(t)+\rho_{\Lambda}(t)
    \end{equation}
    where $t$ represents the cosmic time and the indices $m,r,\Lambda$ stand for matter, radiation and cosmological constant. For each component the equation of state (\ref{lic14}) remains valid.

    In what follows, we will briefly discuss each fluid component and we will try to point out it's contribution to the total density (which influences directly the time evolution of the Universe).

    \paragraph{Matter.} Besides the ordinary luminous matter (protons, neutrons, electrons etc.) it seems that the Universe contains also some kind of exotic non-barionic dark matter \cite{67}, which lies beyond the Standard Model of elementary particles. From indirect cosmological observations (CMB, rotations curves etc) we conclude that the Universe contains much more dark matter than ordinary luminous one.

    The total matter density (at a given time) can be written as the sum of barionic matter $(b)$ and non-barionic dark matter $(dm)$
    \begin{equation}\label{lic17}
    \rho_m(t)=\rho_b(t)+\rho_{bm}(t)
    \end{equation}
    In what follows, we will assume that the thermic energy of the matter particles is negligible compared with its rest mass and as a consequences we can consider matter to be pressureless (i.e. dust). If this is the case, then the matter will have a parameter of state $w=0$. Taking this into account equation (\ref{lic15}) will give for the time evolution of matter density the following expression
    \begin{equation}\label{lic18}
    \rho_m(t)=\rho_{m,0}\left(\frac{a_0}{a(t)}\right)^3
    \end{equation}
    where $ \rho_m(t_0)\equiv \rho_{m,0} $ represents the matter density at present time $t_0$.

    \paragraph{Radiation.} Includes also, besides photons, other species of particles with rest mass zero or almost zero. A typical example are neutrinos, which have a rest mass very close to zero. For radiation the total mass density can be written as
    \begin{equation}\label{lic19}
    \rho_r(t)=\rho_{\gamma}(t)+\rho_{\nu}(t)
    \end{equation}
    It can be showed \cite{64} that for radiation the state parameter $w$ is equal with $1/3$. At any moment of time the radiation density can be written as
    \begin{equation}\label{lic20}
    \rho_r(t)=\rho_{r,0}\left(\frac{a_0}{a(t)}\right)^4
    \end{equation}
    \paragraph{Cosmological constant.} The empty space can be also viewed as a perfect fluid, which obeys the equation of state $p=-\rho$. From this equation of state we can conclude that the pressure of empty space is negative, thus the state parameter in this case would be $w=-1$. In this case equation (\ref{lic15}) will give us
    \begin{equation}\label{lic21}
    \rho_{\Lambda}=\rho_{\Lambda,0}=\frac{\Lambda}{8\pi G}
    \end{equation}
    from which we see that the energy density of empty space is the same at any time.

    Putting together relations (\ref{lic18}), (\ref{lic20}) and (\ref{lic21}) we obtain the following time evolution of the total mass density
    \begin{equation}\label{lic22}
    \rho(t)=\rho_{m,0}\left(\frac{a_0}{a(t)}\right)^3+\rho_{r,0}\left(\frac{a_0}{a(t)}\right)^4 +\rho_{\Lambda,0}\ ,
    \end{equation}
    From (\ref{lic22}) we can observe that the contribution to the total mass density of different cosmological fluid components varies during the history of the Universe, being sensitive to the present values $\rho_{m,0}, \rho_{r,0}, \rho_{\Lambda,0} $. Relation (\ref{lic22}) also tell us that at the beginning radiation was dominant over the other two components and as the Universe evolves the matter starts to dominate over radiation. However, as time goose by in the end the cosmological constant will take over and becomes the dominant term in the total mass density.

    \subsection{Cosmological Parameters}

    In the previous section we wrote the time evolution for the total mass density of matter contained in the Universe, which is important for determining the evolution of the scale factor $a(t)$. Thus, this evolution can be obtained (in our simplified model) only by giving some few cosmological parameters at a particular moment of time $t_*$, which is usually the time at the present epoch $t_0$. Thus, our cosmological model is completely specified by the three values of the following quantities

    \[ \rho_{m,0} \ \ \ \ \ \rho_{r,0} \ \ \ \ \ \rho_{\Lambda,0} \]

    Instead of the above parameters, most of the times it is more useful to use dimensionless quantities, known under the name of density parameters, defined in the following way
    \begin{equation}\label{lic23}
    \Omega_i(t)\equiv\displaystyle\frac{8\pi G}{3H^2(t)}\rho_i(t)
    \end{equation}
    where $i$ stands for $m,r$ or $\Lambda$ and $H(t)$ is the Hubble parameter defined by
    \begin{equation}\label{lic24}
    H(t)=\frac{\dot a(t)}{a(t)}
    \end{equation}
    Thus, a cosmological model can be completely specified by giving the four dimensionless parameters at some particular time (usually the present time)
    \begin{equation}\label{lic25}
    H_0 \ \ \ \ \ \Omega_{m,0} \ \ \ \ \ \Omega_{r,0} \ \ \ \ \ \Omega_{\Lambda,0}
    \end{equation}
    According to the latest results from the Plank mission \cite{9} the parameters $\Omega$ have the following values at the present time
    \begin{equation}\label{lic26}
    H_0\approx 67.11 Kms^{-1}Mpc^{-1} \ \ \ \ \Omega_{m,0}\approx 0.319 \ \ \ \ \Omega_{r,0}\approx 5\times 10^5 \  \ \ \ \Omega_{\Lambda,0}\approx 0.68
    \end{equation}
    If we introduce the density parameters $\Omega_i$ into second equation from (\ref{lic10}) we have
    \begin{equation}\label{lic27}
    1=\Omega_m+\Omega_r+\Omega_{\Lambda}-\displaystyle\frac{k}{H^2a^2}
    \end{equation}
    By introducing the curvature density parameter
    \begin{equation}\label{lic28}
    \Omega_k(t)=-\displaystyle\frac{k}{H^2(t)a^2(t)}
    \end{equation}
    then the cosmological density parameter will obey, at any time, the following simple equation (which can be seen also as a constraint)
    \begin{equation}\label{lic29}
    1=\Omega_m+\Omega_r+\Omega_{\Lambda}+\Omega_k
    \end{equation}
    From (\ref{lic29}) it can be seen that if we know the sum of the first three terms, then we can determine from it the spatial curvature of the Universe. There are three possibilities

    \ \

    $ \Omega_m+\Omega_r+\Omega_{\Lambda}<1 \longleftrightarrow  $ negative spatial curvature $ (k=-1) \longleftrightarrow $ closed Universe

    $ \Omega_m+\Omega_r+\Omega_{\Lambda}=1  \longleftrightarrow $ zero spatial curvature  $ (k=0) \longleftrightarrow $ flat Universe

    $ \Omega_m+\Omega_r+\Omega_{\Lambda}>1 \longleftrightarrow $ positive spatial curvature $ (k=+1) \longleftrightarrow $ open Universe

    \ \

    Although the density parameters $ \Omega_m, \Omega_r, \Omega_{\Lambda}$ are, in general, time dependent, their sum dose not change sign so that the Universe can not pass from a certain FRW geometry (specified by the spatial curvature) to another one.

    We can also introduce a total density parameter
    \begin{equation}\label{lic30}
    \Omega=\Omega_m+\Omega_r+\Omega_{\Lambda}=1-\Omega_k
    \end{equation}
    related to the total mass density (\ref{lic16}) by
    \begin{equation}\label{lic31}
    \Omega=\displaystyle\frac{8\pi G}{3H^2}\rho .
    \end{equation}
    We see from (\ref{lic29}) and (\ref{lic30}) that in order for the Universe to be a flat one we must impose the condition $\Omega=1$, which together with (\ref{lic31}) leads us to the definition of a critical density given (at the present epoch) by
    \begin{equation}\label{lic32}
    \rho_{crit}\equiv\displaystyle\frac{3H_0^2}{8\pi G}
    \end{equation}

    \subsection{Time evolution of the scale factor}

    The second Friedman equation (\ref{lic10})
    \begin{equation}\label{lic33}
    H^2=\frac{8\pi G}{3}(\sum_{i} \rho_i)-\frac{k}{a^2}
    \end{equation}
    can be rewritten, taking into account the cosmological parameters introduced in the previous section, as
    \begin{equation}\label{lic34}
    H^2=H^2_0(\Omega_{m,0}a^{-3}+\Omega_{r,0}a^{-4}+\Omega_{\Lambda,0}+\Omega_{k,0}a^{-3}) .
    \end{equation}
    Now, let us also rewrite the first Friedman equation (\ref{lic10}) as a function of cosmological parameters. In order to do that, we first rearrange the equation as
    \begin{equation}\label{lic35}
    -\displaystyle\frac{a\ddot a}{\dot a^2}=\displaystyle\frac{8\pi G}{3H^2}\sum_{i} \rho_i(1+3w_i) .
    \end{equation}
    and after giving values to $w$ and introducing the densities $\Omega$ from (\ref{lic23}) we arrive at the following expression
    \begin{equation}\label{lic36}
    q=\frac{1}{2}(\Omega_m+2\Omega_r-2\Omega_{\Lambda})
    \end{equation}
    where $q$ is the so called deceleration parameter, defined by
    \begin{equation}\label{lic37}
    q\equiv-\displaystyle\frac{a(t)\ddot a(t)}{\dot a^2(t)} .
    \end{equation}
    If we introduce the definition of the Hubble parameter (\ref{lic24}) into eq. (\ref{lic34}) we obtain a differential equation for the evolution of the scale factor $a(t)$ at any moment of time $t$
    \begin{equation}\label{lic38}
    \left(\frac{da}{dt}\right)^2=H^2_0(\Omega_{m,0}a^{-3}+\Omega_{r,0}a^{-4}+\Omega_{\Lambda,0}a^2+1-\Omega_{m,0}-\Omega_{r,0}-\Omega_{\Lambda,0}),
    \end{equation}
    We observe from the above equation that the form of the scale factor, and thus the evolution of the Universe, is completely determined by a set of cosmological parameters $(H_0,\Omega_{i,0})$ given at the present epoch (which is quite remarkable). In general eq. (\ref{lic38}) dose not have analytical solutions, case in which a numerical investigation is mandatory. However, there are a few particular cases from which we can deduce some analytical cosmological models and this is what we will do in the next section.

    \section{Analytical cosmological models}

    Depending on the values of the cosmological parameters $\Omega_{m,0}, \Omega_{r,0}$ and $\Omega_{\Lambda,0}$, equation (\ref{lic38}) can be solved analytically. There are two main classes of analytical models, namely: the Friedman models (in which the cosmological constant is absent) and the Lemaitre models (which poses a non-zero $\Lambda$). In what follows, we will briefly discuss these models.

    \subsection{Friedman Models}

    Cosmological models for which we assume that the cosmological constant is zero and in which a matter and radiation density are present, are known under the name of Friedman models. This models obey the equation
    \begin{equation}\label{lic39}
    \ddot a=-\frac{4\pi G}{3}(1+3w)\rho a
    \end{equation}
    If we assume that $1+3w$ remains always positive, then the 'acceleration' $\ddot a/a$ will result to be negative conform (\ref{lic39}). Because $a(t_0)>0$ (by definition) and $H(t_0)>0$ (we observe the red shift of galaxies) the curve $a(t)$ will be concave towards the $t-$axis (see fig. 1). From the figure can be seen that the curve $a(t)$ intersects the $t-$axis at a point which is more close to the present time $t_0$ than the time $t$ at which the tangent, that passes through point $(t_0,a(t_0))$, intersects the $t-$axis. We refer at the time when $a(t)$ intersects the $t-$axis as being $t=0$. Thus, we see that at a finite moment of time in the past, the following condition is fulfilled
    \begin{equation}\label{lic40}
    a(0)=0
    \end{equation}
    \begin{figure}[h]
    \centering
    \includegraphics[scale=0.6]{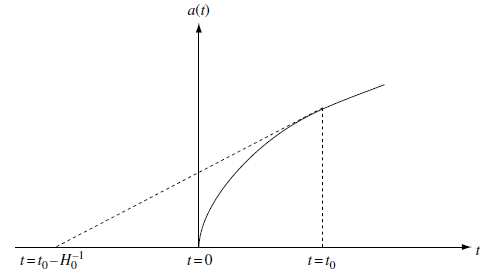}
    \caption{\footnotesize Diagram that illustrates the fact that the age of the Universe, for all Friedman models, is smaller than the Hubble time $H_0^{-1}$}
    \end{figure}
    The point at which $t=0$ can be seen as the beginning of the Universe. The time passed from the moment at which the tangent to the curve $a(t)$ meets the $t-$axis is given by \cite{72}
    \begin{equation}\label{lic41}
    \frac{1}{H_0}=\frac{a(t_0)}{\dot a(t_0)}
    \end{equation}
    We known that for $0<t<t_0$ we have $\ddot a<0$, thus we can say that the real age of the Universe is smaller than the Hubble time
    \begin{equation}\label{lic42}
    t_0<H_0^{-1}
    \end{equation}
    From the above discussion we can conclude that all Friedman models present a Big-Bang moment at a finite point in the past. It can be shown \cite{72}, \cite{73} that the evolution of the scale factor $a(t)$ around the point $t=0$ is independent of the spatial curvature (given by the sign of $k$). However, the future evolution of the Universe depends crucially on the spatial curvature. There are three qualitatively scenarios for the evolution of the Universe, depending on the value of $\Omega_{k,0}$

    $ \Omega_{k,0}<0 \leftrightarrow $  closed Universe $ (k=-1) \leftrightarrow \lim_{a\to\infty}\dot a= $const

    $ \Omega_{k,0}=0 \leftrightarrow $  flat Universe  $ (k=0) \leftrightarrow \lim_{a\to\infty}\dot a=0 $

    $ \Omega_{k,0}>0 \leftrightarrow $  open Universe $ (k=+1) \leftrightarrow \lim_{a\to a_{max}}\dot a=0 $

    One of the main feature of the Friedman models is represented by the fact that dynamical evolution of the models is closely related to what type of geometry the Universe has (fig. 2)
    \begin{figure}[h]
    \centering
    \includegraphics[scale=0.6]{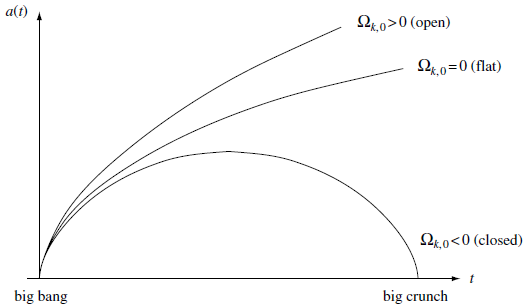}
    \caption{\footnotesize Evolution of the scale factor for the Friedman models}
    \end{figure}
    For the reminder of this paragraph we will discuss some particular Friedman models (for more details see \cite{1}).

    \paragraph{'dust-only' Friedman models $ (\Omega_{\Lambda,0}=0,\Omega_{r,0}=0) $} In this case eq. (\ref{lic38}) becomes
    \begin{equation}\label{lic43}
    \dot a^2=H^2_0(\Omega_{m,0}a^{-1}+1-\Omega_{m,0})
    \end{equation}
    from which we obtain $t$ as
    \begin{equation}\label{lic44}
    t=\frac{1}{H_0}\int^a_0 \left[\displaystyle\frac{x}{\Omega_{m,0}+(1-\Omega_{m,0})x}\right]^{\frac{1}{2}}\mathrm{d}x
    \end{equation}
    After performing the integration for each of the cases $ \Omega_{m,0}<1,\Omega_{m,0}=1 $ and $ \Omega_{m,0}>1 $, we obtain the following solutions for $a(t)$

    $\bullet$ For $ \Omega_{m,0}<1$ ($k=-1$) the integral (\ref{lic44}) can be performed with the help of the substitution
    \begin{equation}\label{lic45}
    x=\displaystyle\frac{\Omega_{m,0}}{\Omega_{m,0}-1}\sinh^2\frac{\Psi}{2}
    \end{equation}
    where $\Psi$ is known under the name of development angle, taking values between $0$ and $\pi$. In this case we obtain the following parametric equations for the evolution of $a(t)$
    \begin{equation}\label{lic46}
    \begin{split}
    &a=\displaystyle\frac{\Omega_{m,0}}{2(1-\Omega_{m,0})}(\cosh\Psi-1) \\
    &t=\displaystyle\frac{\Omega_{m,0}}{2H_0(1-\Omega_{m,0})^{\frac{3}{2}}}(\sinh\Psi-\Psi)
    \end{split}
    \end{equation}
    $\bullet$ The case $\Omega_{m,0}=1$ ($k=0$) integrates immediately, giving us the solution
    \begin{equation}\label{lic47}
    a(t)=\left(\frac{3}{2}H_0t\right)^{\frac{2}{3}} .
    \end{equation}
    also known under the name of Einstein-de-Sitter model.

    $\bullet$ The case $\Omega_{m.0}>1$ ($k=+1$) can be solved with the help of a similar substitution as in (\ref{lic45}) if we replace the hyperbolic function $\sinh$ with the $\sin$ function, thus obtaining in the end a cycloid form for the curve $a(t)$, given by the parametric equations
    \begin{equation}\label{lic48}
    \begin{split}
    &a=\displaystyle\frac{\Omega_{m,0}}{2(\Omega_{m,0}-1)}(1-\cos\Psi) \\
    &t=\displaystyle\frac{\Omega_{m,0}}{2H_0(\Omega_{m,0}-1)^{\frac{3}{2}}}(\Psi-\sin\Psi)
    \end{split}
    \end{equation}

    \paragraph{'radiation-only' Friedman models $ (\Omega_{\Lambda,0}=0,\Omega_{m,0}=0) $} can be obtained from the equation
    \begin{equation}\label{lic49}
    \dot a^2=H^2_0(\Omega_{r,0}a^{-2}+1-\Omega_{r,0})
    \end{equation}
    from which $t$ is equal to
    \begin{equation}\label{lic50}
    t=\frac{1}{H_0}\int^a_0\displaystyle\frac{x}{\sqrt{\Omega_{r,0}+(1-\Omega_{r,0})x^2}}\mathrm{d}x
    \end{equation}
    Solving eq. (\ref{lic49}) we obtain two different solutions

    $\bullet$ For $\Omega_{r,0}=1$ ($k=0$)
    \begin{equation}\label{lic51}
    a(t)=(2H_0t)^{\frac{1}{2}}
    \end{equation}
    $\bullet$ If $\Omega_{r,0}<1$ ($k=-1$) or $\Omega_{r,0}>1$ ($k=+1$) after integration of (\ref{lic49}) we get
    \begin{equation}\label{lic52}
    a(t)=\left(2H_0\Omega_{r,0}^{\frac{1}{2}}\right)^{\frac{1}{2}}\left(1+\displaystyle\frac{1-\Omega_{r,0}}{2\Omega_{r,0}^{\frac{1}{2}}H_0t}\right)^{\frac{1}{2}}
    \end{equation}
    For all Friedman models analysed so far we can also calculate the quantities $\rho_{m/r}(t),H(t),\Omega_{m/r}(t),\Omega(t) $ with the help of relations (\ref{lic18}), (\ref{lic20}), (\ref{lic24}), (\ref{lic30}).

    \paragraph{'spatially-flat' Friedman models $ (\Omega_{\Lambda,0}=0,\Omega_{m,0}+\Omega_{r,0}=1) $ }In this case eq. (\ref{lic38}) becomes
    \begin{equation}\label{lic53}
    \dot a^2=H^2_0(\Omega_{m,0}a^{-1}++1-\Omega_{r,0}a^{-2})
    \end{equation}
    with $t$ given by
    \begin{equation}\label{lic54}
    t=\frac{1}{H_0}\int^a_0\displaystyle\frac{x}{\sqrt{\Omega_{m,0}x+\Omega_{r,0}}}\mathrm{d}x
    \end{equation}
    If we make the transformation of variables as $ y=\Omega_{m,0}x+\Omega_{r,0} $ the integral (\ref{lic54}) becomes
    \begin{equation}\label{lic55}
    H_0t=\displaystyle\frac{2}{3\Omega_{m,0}^2}\left[(\Omega_{m,0}a+\Omega_{r,0})^{\frac{1}{2}}(\Omega_{m,0}a- 2\Omega_{r,0})+2\Omega_{m,0}^{\frac{3}{2}}\right].
    \end{equation}
    from which we can extract a messy formula for $a(t)$.

    \subsection{Lemaitre models}

    The models in which a cosmological constant $\Lambda$ is present are known under the name of Lemaitre cosmological models. There exists different types of Lemaitre models differentiated by the values of the cosmological parameters $ \Omega_{m,0}, \Omega_{r,0}$ and $\Omega_{\Lambda,0}$. In this review we will briefly discuss only "matter" Lemaitre models for which $\Omega_{r,0}=0$. The other models that contain only radiation or both radiation and matter are very similar with the "matter" lemaitre models.

    \paragraph{'matter-only' Lemaitre models with an arbitrary spatial curvature $ (\Omega_{r,0}=0) $ \cite{1}} For this models equation (\ref{lic38}) can be written as
    \begin{equation}\label{lic56}
    \dot a^2=H^2_0(\Omega_{m,0}a^{-1}+\Omega_{\Lambda,0}+\Omega_{k,0})
    \end{equation}
    The solutions of this equation are complicated, being expressed in terms of elliptical functions \cite{74}. For small values of $t$ an approximative solution of eq. (\ref{lic56}) is given by
    \begin{equation}\label{lic57}
    a(t)=\left(\frac{3}{2}H_0\sqrt{\Omega_{m,0}t}\right)^{\frac{2}{3}}
    \end{equation}
    As the Universe continues to expand, the energy-mass density is decreasing and thus the cosmological constant starts to take over and becomes dominant to the total energy density. Thus, the asymptotic solution (for large $t$) of eq. (\ref{lic56}) will be
    \begin{equation}\label{lic58}
    a(t)\propto e^{H_0\sqrt{\Omega_{\Lambda,0}}t}
    \end{equation}
    From eq. (\ref{lic57}) we can deduce that after the initial Big-Bang the expansion of the Universe is decelerating. However, eq. (\ref{lic58}) also tell us that the expansion is accelerating for large values of $t$. In order for the two evolutions to be possible the Universe must have a transition period during which the second time-derivative of the scale factor vanishes ($\ddot a=0$). Differentiating eq. (\ref{lic56}) we obtain
    \begin{equation}\label{lic59}
    \ddot a=\frac{1}{2}H_0^2(2\Omega_{\Lambda,0}a-\Omega_{m,0}a^{-2})
    \end{equation}
    from which (using $\ddot a=0$) we get the transition point
    \begin{equation}\label{lic60}
    a_{*}=\left(\displaystyle\frac{\Omega_{m,0}}{2\Omega_{\Lambda,0}}\right)^{\frac{1}{3}}
    \end{equation}
    In the vicinity of this transition point it is even possible to obtain an analytical expression for the evolution of the scale factor. This can be done in the following way: first we make a Taylor expansion around the point $t=t_{*}$ for
    \begin{equation}\label{lic61}
    \begin{split}
    &a(t)\approx a_{*}(t)+\dot a_{*}(t)(t-t_{*}) \\
    &\dot a^2(t)\approx \dot a^2_{*}(t)+\ddot a_{*}(t)\dddot a_{*}(t)(t-t_{*})^2
    \end{split}
    \end{equation}
    where $t_{*}$ it the time corresponding to $a_{*}$. Then we eliminate $t-t_{*}$ between the two relations in (\ref{lic61}) to get
    \begin{equation}\label{lic62}
    \dot a^2\approx \dot a^2_{*}+\frac{\dddot a_{*}(a-a_{*})^2}{\dot a_{*}}
    \end{equation}
    By introducing this result into relation (\ref{lic59}) and then back in eq. (\ref{lic56}) we arrive at the following equation
    \begin{equation}\label{lic63}
    \dot a^2\approx H^2_0[\Omega_{k,0}+3\Omega_{\Lambda,0}a^2_{*}+3\Omega_{\Lambda,0}(a-a_{*})^2]
    \end{equation}
    which is easy to integrate and to obtain the next solution
    \begin{equation}\label{lic64}
    a(t)=a_{*}+a_{*}\left[1+ \frac{1}{3}\Omega_{k,0}\left(\frac{1}{4}\Omega_{\Lambda,0}\Omega_{m,0}^2\right)^{-\frac{1}{3}}\right]^{\frac{1}{2}} \sinh\left[H_0(3\Omega_{\Lambda,0})^{\frac{1}{2}}(t-t_{*})\right]
    \end{equation}
    An interesting property of this model is the existence of an almost plat region in the vicinity of the point $a_{*}$ (fig.3)
    This happens only if the spatial curvature is positive ($k=+1$) \cite{1}.
    \begin{figure}[h]
    \centering
    \includegraphics[scale=0.6]{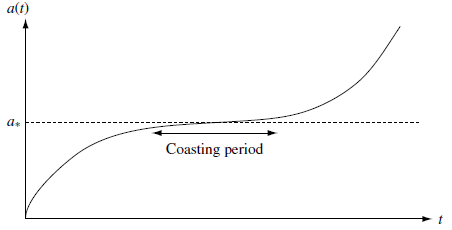}
    \caption{\footnotesize Shape of the scale factor for a Lemaitre model with $k=+1$}
    \end{figure}
    This flat region can be made almost as big as one wishes to by imposing the condition
    \begin{equation}\label{lic65}
    \frac{1}{3}\Omega_{k,0}\left(\frac{1}{4}\Omega_{\Lambda,0}\Omega_{m,0}^2\right)^{-\frac{1}{3}} \rightarrow -1
    \end{equation}

    \paragraph{'spatially-flat matter-only' Lemaitre models $ (\Omega_{r,0}=0,\Omega_{m,0}+\Omega_{\Lambda,0}=1) $} This models have an exact analytical solution. This solution can be obtain from the equation
    \begin{equation}\label{66}
    \dot a^2=H_0^2[(1-\Omega_{\Lambda,0})a^{-1}+\Omega_{\Lambda,0}a^2]
    \end{equation}
    from which
    \begin{equation}\label{lic67}
    t=\frac{1}{H_0}\int_0^a\displaystyle\frac{x}{\sqrt{(1-\Omega_{\Lambda,0})+\Omega_{\Lambda,0}x^4}}\mathrm{d}x
    \end{equation}
    By making the following change of variables
    \begin{equation}\label{lic68}
    y^2=\displaystyle\frac{|\Omega_{\Lambda,0}|}{1-\Omega_{\Lambda,0}}x^3
    \end{equation}
    the integral (\ref{lic67}) becomes
    \begin{equation}\label{lic69}
    H_0t=\displaystyle\frac{2}{3\sqrt{|\Omega_{\Lambda,0}|}}\displaystyle\int_0^{\sqrt{a^3\displaystyle \frac{|\Omega_{\Lambda,0}|}{1-\Omega_{\Lambda,0}}}}\displaystyle\frac{\mathrm{d}y}{\sqrt{1\pm y^2}}
    \end{equation}
    from which we obtain the solution
   \begin{equation}\label{lic70}
    \begin{gathered}
    H_0t=\displaystyle\frac{2}{3\sqrt{|\Omega_{\Lambda,0}}} \left\{ \begin {array}{ll}
        \rm arcsinh\sqrt{a^3\displaystyle\frac{|\Omega_{\Lambda,0}|}{1-\Omega_{\Lambda,0}}} & \mbox{if  $ \Omega_{\Lambda,0}>0 $} \\
        \arcsin\sqrt{a^3\displaystyle\frac{|\Omega_{\Lambda,0}|}{1-\Omega_{\Lambda,0}}} & \mbox{if  $ \Omega_{\Lambda,0}<0 $}\end{array} \right.
    \end{gathered}
    \end{equation}

    \subsection{De Sitter model}

    This model is a very particular case of spatially-flat Lemaitre ($k=0$) for which the Universe undergoes an exponential expansion. The model is specified by the following particular values of the cosmological parameters
    \begin{equation}\label{lic71}
    \Omega_{m,0}=0, \ \ \ \Omega_{r,0}=0, \ \ \ \Omega_{\Lambda,0}=1
    \end{equation}
    In the case of de Sitter the equation (\ref{lic38}) for $a(t)$ becomes
    \begin{equation}\label{lic72}
    \left(\frac{\dot a}{a}\right)^2=H_0^2
    \end{equation}
    Thus we observe that the Hubble parameter $H(t)$ is constant in time. Solving (\ref{lic72}) we obtain the de Sitter solution
    \begin{equation}\label{lic73}
    a(t)=e^{H_0(t-t_0)}=e^{\sqrt{\frac{\Lambda}{3c}}(t-t_0)}
    \end{equation}
    It is interesting to note that in de Sitter model we don't have an initial Big-Bang singularity at some finite time in the past.

\section{Metric f(R) cosmology}

    Among the motivations for studying $f(R)$ theories of gravity we also identify the need to explain the apparent late-time acceleration of the Universe \cite{39}, \cite{40}. On large scales the Universe is considered to have the proprieties of homogeneity and isotropy. Then irrespective of the theory of gravity used, the main assumptions of cosmology remain the same since they are related only to the symmetries that Universe poses. Once a gravity theory is chosen, one can insert the FRW ansatz
    \begin{equation}\label{cf1}
    ds^2=-dt^2 + a(t)^2 \left[ \frac{dr^2}{1-kr^2}+r^2\left( d\theta^2+\sin^2\!\theta\, d\varphi^2 \right) \right]
    \end{equation}
    together with the stress energy tensor of a perfect fluid form
    \begin{equation}\label{cf2}
    T^{\mu\nu} = \left( \rho + p \right)u^\mu u^\nu + g^{\mu\nu} p.
    \end{equation}
    We can now derive the generalised Friedman equations for metric $f(R)$ gravity. We introduce the above formulae in (\ref{3.8}) together with the expression for the Ricci scalar
    \begin{equation}\label{cf3}
    R=6\left\{ \frac{\ddot a}{a}+\left(\frac{\dot a}{a} \right)^2 + \frac{k}{a^2}\right\}=6\left( \dot H+2H^2+\frac{k}{a^2} \right)
    \end{equation}
    where $H=\dot a /a $ represents the Hubble parameter. We obtain for the time-time and space-space components of the equation (\ref{3.8}) the following expressions
    \begin{equation}\label{cf4}
    \begin{split}
    &\left(\frac{\dot a}{a} \right)^2 + \frac{k}{a^2} - \frac{1}{3f'}\left\{ \frac{1}{2}\left(f-Rf'\right) -3\frac{\dot a}{a}\dot Rf'' \right\}=\frac{1}{3}\kappa\rho \\
    &2\frac{\ddot a}{a}+\left(\frac{\dot a}{a} \right)^2 + \frac{k}{a^2}+ \frac{1}{f'}\left\{ 2\left( \frac{\dot a}{a} \right)\dot R f'' + \ddot R f'' + \dot R^2f'''- \frac{1}{2}\left(f-Rf'\right) \right\}= -\kappa p
    \end{split}
    \end{equation}
    We can introduce a curvature pressure and density given by \cite{54}
    \begin{equation}\label{cf5}
    \begin{split}
    &p_{curv} = \frac{1}{\kappa f'}\left\{ 2\left( \frac{\dot a}{a} \right)\dot Rf'' + \ddot Rf'' + \dot R^2f'''- \frac{1}{2}\left(f-Rf'\right) \right\} \\
    &\rho_{curv} = \frac{1}{\kappa f'}\left\{ \frac{1}{2}\left(f-Rf'\right) -3\frac{\dot a}{a}\dot Rf'' \right\}
    \end{split}
    \end{equation}
    and if we define a total pressure and energy density as
    \begin{equation}\label{cf6}
    \begin{split}
    & p_{tot} = p_{curv} +p \\
    & \rho_{tot} = \rho_{curv} + \rho
    \end{split}
    \end{equation}
    then we can rewrite the Friedman equations as
    \begin{equation}\label{cf6}
    \left(\frac{\dot a}{a} \right)^2 + \frac{k}{a^2}=\frac{1}{3}\kappa\rho_{tot}
    \end{equation}
    and
    \begin{equation}\label{cf7}
    2\frac{\ddot a}{a}+\left(\frac{\dot a}{a} \right)^2 + \frac{k}{a^2}=-\kappa p_{tot}
    \end{equation}
    Combining eq (\ref{cf6}) and (\ref{cf7}), we obtain
    \begin{equation}\label{cf8}
    \frac{\ddot a}{a}=-\frac{1}{6}\left( \rho_{tot}+3p_{tot} \right)
    \end{equation}
    We see that the accelerated behaviour is achieved if
    \begin{equation}\label{cf9}
    \rho_{tot} + 3p_{tot} < 0
    \end{equation}
    which means
    \begin{equation}\label{cf10}
    \rho_{curv} > \frac{1}{3}\rho_{tot}
    \end{equation}
    assuming that all matter components have non-negative pressure.

    We can  define a curvature equation of state parameter $ w_{curv} $ of modified gravity which can be expressed as
    \begin{equation}\label{cf11}
    w_{curv}\equiv\frac{p_{curv}}{\rho_{curv}}=\frac{ 2\left( \frac{\dot a}{a} \right)\dot Rf'' + \ddot Rf'' + \dot R^2f'''- \frac{1}{2}\left(f-Rf'\right)}{ \frac{1}{2}\left(f-Rf'\right) -3\frac{\dot a}{a}\dot Rf''}
    \end{equation}
    The asymptotic de Sitter spacetime is obtained in for $ w_{curv}=-1 $, case in which the following condition must hold for any $f(R)$ model
    \begin{equation}\label{cf12}
    \frac{f'''}{f''} = \frac{\dot R\left( \frac{\dot a}{a} \right) - \ddot R}{\dot R^2}
    \end{equation}
     The form of $ f(R) $ function is the main ingredient to obtain a curvature quintessence \cite{57} (the term used for an accelerated expansion without a need of a scalar field).

\subsection{$1/R$ cosmology}

    In what follows we will develop a model that could explain the late-time acceleration. Among the first attempts that tried to explain this was the work of Capozziello \cite{57}. Here we will consider the model proposed by Carroll et all. in \cite{55}. Let us stress out form the beginning that this is more of a toy model because it has been showed that this model present violent instability in the matter sector \cite{67}, in fact there is no matter dominated era \cite{68}, \cite{69}. A proposal to generalise this model and to avoid some of the above problems was made in \cite{79}. In the present model we choose the function $f(R)$ to be of the form
    \begin{equation}\label{cf17}
    f(R)=R-\frac{\mu^4}{R}
    \end{equation}
    Introducing this in eq. (\ref{3.8}) we obtain
    \begin{equation}\label{cf18}
    \left( 1+\frac{\mu^4}{R^2}\right)R_{\mu\nu}-\frac{1}{2}\left(1-\frac{\mu^4}{R}\right)R g_{\mu\nu} + (g_{\mu\nu}\Box-\nabla_\mu\nabla_\nu)\left( 1+\frac{\mu^4}{R^2}\right)=\frac{T_{\mu\nu}}{M_P^2}
    \end{equation}
    where $M_P^2=1/8\pi G$ is the reduced Planck Mass. The trace of eq. (\ref{cf18}) reads
    \begin{equation}\label{cf19}
    3\Box F(R)+RF(R)-2f(R)=\frac{T}{M_P^2}
    \end{equation}
    If we are interested in finding constant-curvature vacuum solutions, for which $\nabla_\mu R=0$, we need to solve the simplified version of eq. (\ref{cf19})
    \begin{equation}\label{cf20}
    RF(R)-2f(R)=0
    \end{equation}
    Now if we substitute (\ref{cf17}) in eq. (\ref{cf20}) we obtain
    \begin{equation}\label{cf21}
    \left( 1+\frac{\mu^4}{R^2}\right)R-2\left(R-\frac{\mu^4}{R}\right)=0
    \end{equation}
    which can be solved immediately to give
    \begin{equation}\label{cf22}
    R=\pm\sqrt{3}\mu^2
    \end{equation}
    Thus, we find the interesting result that the constant-curvature vacuum solutions are not Minkowski space, but rather are de Sitter space and anti de Sitter space \cite{55}.

    The time-time component of the field equation (\ref{cf18}) for the flat ($k=0$) FRW metric (\ref{cf1}) is
    \begin{equation}\label{cf23}
    3H^2-\frac{\mu^4}{12R^3}\left( 2H\ddot H+15H^2\dot H+2\dot H^2+6H^4 \right)=\frac{\rho}{M_P^2}
    \end{equation}

    The other independent equation is the space-space component of eq. (\ref{cf18})
    \begin{equation}\label{cf24}
    3H^2+2\dot H-\frac{\mu^4}{36R^2}\left( 4H\dot H+ 9H^2 + 2\frac{\ddot R}{R}- 6\frac{\dot R^2}{R^2} + 4H\frac{\dot R}{R}\right)=-\frac{p}{M_P^2}
    \end{equation}
    By setting $\mu=0$ we recover the usual Friedman equations
    \begin{equation}\label{cf25}
    \begin{split}
    &3H^2=\frac{\rho}{M_P^2}\\
    &2\dot H + 3H^2=-\frac{p}{M_P^2}
    \end{split}
    \end{equation}
    The new Friedman equations (\ref{cf23}) and (\ref{cf24}) are of forth order and very difficult to solve. Because of this it is more convenient to transform the equations to the Einstain frame. In our case the conformal factor (\ref{c8}) will be
    \begin{equation}\label{cf26}
    \Omega^2=1+\frac{\mu^4}{R^2}=exp\left((\sqrt{\frac{2}{3}}\frac{\phi}{M_P^2}\right)
    \end{equation}
    Next, we would like to find a relation between the conformal transformed Hubble parameter $\tilde H$ and the old one $H$. In order to to that let us write first the following transformation relations
    \begin{equation}\label{cf27}
    \begin{split}
    &d\tilde t=\sqrt{F}dt\\
    &\tilde a=a\sqrt{F}
    \end{split}
    \end{equation}
    Then we can write for the Hubble parameter the followings
    \begin{equation}\label{cf28}
    \begin{split}
    \tilde H=\frac{\tilde a'}{\tilde a}&=\frac{1}{a\sqrt{F}}\left( \dot a+ \frac{a}{2\sqrt{F}}F' \right)\\
    &=\frac{H}{\sqrt{F}}+\frac{1}{2F}\sqrt{\frac{2}{3}}\frac{1}{M_P^2}\exp\left((\sqrt{\frac{2}{3}}\frac{\phi}{M_P^2}\right)\phi'\\
    &=\frac{H}{\sqrt{F}}+\frac{\phi'}{\sqrt{6}M_P^2}
    \end{split}
    \end{equation}
    where a prime denotes differentiations with respect to conformal time and we also used the following formula
    \begin{equation}\label{cf29}
    \frac{d}{d\tilde t}=\frac{dt}{d\tilde t}\frac{d}{dt}=\frac{\sqrt{F}}{dt}
    \end{equation}
    The scalar field potential defined in (\ref{c15}) now reads
    \begin{equation}\label{cf30}
    V(\phi)=\mu^2M_P^2\frac{\sqrt{F-1}}{F^2}
    \end{equation}

    Note that the scalar field potential is zero both when the scalar field itself is zero or when this goes to infinity. The potential achieves it's maximum value at the point where $dV/dF=0$, i.e. at $F=4/3$
    \begin{equation}\label{cf31}
    V(\phi)_{max}=V(\phi)_{F=4/3}=\frac{9\mu^2M_P^2}{16\sqrt{3}}
    \end{equation}
    It is also convenient to define an Einstein frame matter energy-momentum tensor by
    \begin{equation}\label{cf32}
    \tilde T_{\mu\nu}=(\tilde\rho+\tilde p)\tilde u_{\mu}\tilde u_{\nu}+\tilde p\tilde g_{\mu\nu}
    \end{equation}
    If we compare (\ref{cf32}) with (\ref{c25}) we can make the following identifications
    \begin{equation}\label{cf33}
    u_{\mu}=\frac{1}{\sqrt{F}}\tilde u_{\mu}, \ \ \ \ \ \rho=F^2\tilde\rho, \ \ \ \ \ p=F^2\tilde p
    \end{equation}
    The cosmological equation of motion in the Einstein frame are given by
    \begin{equation}\label{cf34}
    3\tilde H^2=\frac{1}{M_P^2}\left( \rho_{\phi}+\tilde\rho\right)
    \end{equation}
    \begin{equation}\label{cf35}
    \phi''+3\tilde H\phi'+\frac{dV}{d\phi}-\frac{1-3w}{\sqrt{6}M_P^2}\tilde\rho=0
    \end{equation}
    where
    \begin{equation}\label{cf36}
    \tilde\rho=\frac{C}{\tilde a^{3(1+w)}}\exp{\left(-\frac{1-3w}{\sqrt{6}}\frac{\phi}{M_P}\right)}
    \end{equation}
    \begin{equation}\label{cf37}
    \rho_{\phi}=\frac{1}{2}\phi'^2+V(\phi)
    \end{equation}
    Equation (\ref{cf36}) is obtained from eq. (\ref{lic15}) by replacing $\rho$ and $a$ with their Einstein frame counterparts given in (\ref{cf27}) and (\ref{cf33}). Equation (\ref{cf35}) represents the Klein-Gordon equation of motion for the scalar field given in (\ref{c28}) with $ \tilde T $ given by
    \begin{equation}\label{cf38}
    \tilde T=\tilde\rho-3\tilde p=(1-3w)\tilde\rho
    \end{equation}
    and it can be shown \cite{75} that for $f(R)$ gravity $Q=-1/\sqrt{6}$. The first term in (\ref{c28}) can be expanded as follows
    \begin{equation}\label{cf39}
    \begin{split}
    \Box\phi=&\frac{1}{\sqrt{\tilde g}}\partial_{\mu}\left(\sqrt{\tilde g}\tilde g^{\mu\nu}\partial_{\nu}\phi \right)\\
    &=\frac{\partial_{\mu}\tilde g}{2\tilde g}\tilde g^{\mu\nu}\partial_{\nu}\phi + \partial_{\mu}(\tilde g^{\mu\nu})\partial_{\nu}\phi+\tilde g^{\mu\nu}\partial_{\mu}(\partial_{\nu}\phi) \\
    &=-3\tilde H\phi'-\phi''
    \end{split}
    \end{equation}
    where in the las line we used that $\tilde g=F^4g=\tilde a^6$.

    Let us now make some comments about the solutions of eq. (\ref{cf34}-\ref{cf35}). It is more easy to find vacuum cosmological solutions for which $\rho=p=0$. The beginning of the Universe corresponds to $R\rightarrow \inf$ and $\phi\rightarrow 0$. Depending on what initial conditions we specify for $\phi$ and $\phi '$, there are three qualitatively different solutions \cite{55}.

    (a) Eternal de Sitter: is obtained when the scalar field $\phi$ comes to a rest after reaching the maximum of the potential $V(\phi)$ given in (\ref{cf31}). This state is obtained for a critical value of the initial condition, namely $\phi_i'\equiv\phi_c'$ when the Universe evolves asymptotically to a de Sitter solution which turns out to be very unstable because any perturbation of the field will make it roll away from the maximum of it's potential.

    (b) Power-law acceleration: is achieved for $\phi_i' > \phi_c'$, when the field overshoots the maximum of $V(\phi)$ and the potential can be further approximated by $V(\phi)=\mu^2M_P^2/\sqrt{F}$. In this case it is easily to solve for $ \tilde a(\tilde t)~\tilde t^{4/3}$, which corresponds to $a(t)~t^2$. Thus, the Universe evolves to late-time power-law inflation.

    (c) Future singularity: is obtained when $\phi_i' < \phi_c'$, case in which the scalar field rolls back down to $\phi=0$.

\subsection{$R^2$ cosmology}

    The most representative model of $R^2$ cosmology is the so called Starobinsky \cite{56} model with the help of which inflation can be explained without a need for a scalar field. This model is obtained for
    \begin{equation}\label{c2r1}
    f(R)=R+\alpha R^2
    \end{equation}
    where $\alpha$ is constant and has dimension of inverse mass squared. Now equations (\ref{3.8}) can be written as
    \begin{equation}\label{c2r2}
    \left( 1+2\alpha R^2\right)R_{\mu\nu}-\frac{1}{2}\left(R+\alpha R^2\right)R g_{\mu\nu} + (g_{\mu\nu}\Box-\nabla_\mu\nabla_\nu)\left(1+2\alpha R^2\right)=\frac{T_{\mu\nu}}{M_P^2}
    \end{equation}
    Using eq. (\ref{cf19}) it can be seen immediately that the constant-curvature vacuum solution is the usual Minkowski space-time. After we introduce the FRW anzat in (\ref{c2r2}), the combination of Friedman equations can be written as
    \begin{equation}\label{c2r3}
    \ddot H-\frac{\dot H^2}{2H}+\frac{1}{12\alpha}H=-3H\dot H
    \end{equation}
    \begin{equation}\label{c2r4}
    \ddot R+3H\dot R+\frac{1}{6\alpha}R=0
    \end{equation}
    For the inflationary epoch the first two terms of (\ref{c2r3}) can be neglected relative to others, obtaining
    \begin{equation}\label{c2r5}
    \dot H\approx -\frac{1}{26\alpha}
    \end{equation}
    from which follows immediately the solution
    \begin{equation}\label{c2r6}
    H\approx H_i-\frac{1}{36\alpha}(t-t_i)
    \end{equation}
    and
    \begin{equation}\label{c2r7}
    a\approx a_i\exp\left\{ H_i(t-t_i)-\frac{1}{72\alpha}(t-t_i)^2 \right\}
    \end{equation}
    where $H_i$ and $a_i$ are initial values at the beginning of inflation.

    Like in the previous section we may take a conformal transformation of the form (\ref{c8}) and go to the Einstein frame in order to discuss the inflationary dynamics of the theory. We will choose, for the moment, to ignore all the matter fluids, since they don't play any significant role during the inflationary phase. In the Einstein frame, the field potential (\ref{c15}) reads
    \begin{equation}\label{c2r8}
    \begin{split}
    V(\phi)&=\frac{FR-f}{\kappa^2F^2}=\frac{\alpha}{2\kappa^2}\left(\frac{R}{1+2\alpha R} \right)\\
    &=\frac{1}{8\alpha\kappa^2}\left( 1-e^{-\sqrt{2/3}\kappa\phi} \right)^2
    \end{split}
    \end{equation}
    In the case when $\kappa\phi\gg 1$ it can be seen form (\ref{c2r8}) that the potential is nearly constant, having the value
    \begin{equation}\label{c2r9}
    V(\phi)\approx \frac{1}{8\alpha\kappa^2}
    \end{equation}
    which leads to slow-roll inflation \cite{70}, \cite{71}. In the opposite case, when $\kappa\phi\ll 1$, the potential reduces to
    \begin{equation}\label{c2r10}
    V(\phi)\approx \frac{1}{12\alpha}\phi^2
    \end{equation}
    case in which the field oscillates around $\phi=0$ with a Hubble damping. Using the fact that during inflation the Ricci scalar can be approximated by $R\approx 12H^2-1/6\alpha$ (where we used (\ref{c2r5})), then during inflation holds that $F\approx 24\alpha H^2$. Now, the transformation (\ref{cf27}) gives a relation between the time $\tilde t$ in the Einstein frame and the old time $t$
    \begin{equation}\label{c2r11}
    \tilde t=\int^{t}_{t_i}\!\mathrm{d}t\,\sqrt{F}\approx 2\sqrt{6\alpha}\left\{ H_i(t-t_i)-\frac{1}{72\alpha}(t-t_i)^2 \right\}
    \end{equation}
    Using eq. (\ref{c2r7}) and (\ref{c2r11}), the scale factor $\tilde a=\sqrt{F}a$ evolves as
    \begin{equation}\label{c2r12}
    \tilde a(\tilde t)\approx \left( 1-\frac{1}{72\alpha H_i^2}\frac{\tilde t}{\sqrt{6\alpha}} \right)\tilde a_i e^{\sqrt{6\alpha}\tilde t/2}
    \end{equation}
    where $a_i=2H_ia_i/\sqrt{6\alpha}$. Using (\ref{cf28}), the evolution of the Hubble parameter is given by
    \begin{equation}\label{c2r13}
    \tilde H(\tilde t)\approx \frac{\sqrt{6\alpha}}{2}\left\{1-\frac{1}{36\alpha H_i^2}\left(1-\frac{1}{72\alpha H_i^2}\sqrt{6\alpha}\,\tilde t \right)^{-2} \right\}
    \end{equation}
    We observe that the Hubble parameter is decreasing with time. From (\ref{c2r12}) and (\ref{c2r13}) we can conclude that the Universe expands quasi-exponentially in the Einstein frame.

\section*{Acknowledgements}

This work was supported by a grant of the Romanian National Authority for Scientific Research, Programme for research-Space Technology and Advanced Research-STAR, project nr. 72/29.11.2013 between Romanian Space Agency and West University of Timisoara and in part by ICTP - SEENET-MTP project PRJ-09 Cosmology and Strings. The author would like to thank Professor Ion Cotaescu for reading the manuscript and for useful suggestions and discussions that helped to improve this work. Also special thanks goes to Diana Popescu for reading the manuscript and checking its English.


\begin{thebibliography}{99}


\bibitem{1} M. P. Hobson, G. P. Efstathiou, A. N. Lasenby, "General Relativity, An Introduction for Physicists", Cambridge University Press, 2006.

\bibitem{2} E. Poisson, "A Relativist's Toolkit - The Mathematics of Black-Hole Mechanics", Cambridge University Press, 2004.

\bibitem{3} R.M. Wald, "General Relativity", Chicago University Press, Chicago, 1984.

\bibitem{4} A. Guarnizo, L. Castaneda, J.M. Tejeiro, "Boundary Term in Metric f(R) Gravity: Field Equations in the Metric Formalism", Gen.Rel.Grav. 42 (2010) 2713-2728.

\bibitem{5} T. P. Sotiriou, "Field equations from a surface term", Phys.Rev. D74 (2006) 044016.

\bibitem{6} G. W. Gibbons and S. W. Hawking, "Action Integrals And Partition Functions In Quantum Gravity",
Phys. Rev. D. 15 (1977):2752.

\bibitem{7} S. W. Hawking and G. T. Horowitz, "The Gravitational Hamiltonian, action, entropy and surface
terms", Class. Quant. Grav. 13 (1996):1487-1498.

\bibitem{8} S. Weinberg, "Gravitation and Cosmology", Wiley, 1972.

\bibitem{9} P. A. R. Ade, N. Aghanim, C. Armitage-Caplan, et al., (Planck Collaboration) (31 March 2013). "Planck 2013 Results Papers". Astronomy and Astrophysics. arXiv:1303.5062

\bibitem{10} S. M. Carroll, "The Cosmological constant", Living Rev.Rel. 4 (2001) 1.

\bibitem{11} J. D. Bekenstein, "Relativistic gravitation theory for the MOND paradigm" Phys.Rev. D70 (2004) 083509
Erratum-ibid. D71 (2005) 069901.

\bibitem{12} C. Brans, R. H. Dicke, 1961, "Mach's Principle and a Relativistic Theory of Gravitation", Phys. Rev. 124, 925.

\bibitem{13} G. R. Dvali, G. Gabadadze, M. Porrati, "4-D gravity on a brane in 5-D Minkowski space", Phys.Lett. B485 (2000) 208-214.

\bibitem{14} V. Faraoni, E. Gunzig, P. Nardone, "Conformal transformations in classical gravitational theories and in cosmology", Fund.Cosmic Phys. 20 (1999) 121.

\bibitem{15} V. Faraoni, E. Gunzig, "Einstein frame or Jordan frame?", Int.J.Theor.Phys. 38 (1999) 217-225.

\bibitem{16} R.H. Dicke, "Mach’s Principle and Invariance under Transformation of Units", Phys. Rev.,
125, 2163–2167, (1962).

\bibitem{17} K.-I. Maeda, "Towards the Einstein-Hilbert Action via Conformal Transformation", Phys.
Rev. D, 39, 3159–3162, (1989).

\bibitem{18} D. Wands, "Extended gravity theories and the Einstein–Hilbert action", Class. Quantum
Grav., 11, 269–279, (1994).

\bibitem{19} A. H. Guth, "Inflationary universe: A possible solution to the horizon and flatness problems", 1981, Phys. Rev. D23, 347.

\bibitem{20} G. de Vaucouleurs, Science, 167, 1203 (1970).

\bibitem{21} G. V. Bicknell, "A geometrical Lagrangian for the neutral scalar meson field", Journal of Physics
A Mathematical General, 7:341(345), Feb. 1974.

\bibitem{22} J. D. Barrow, S. Cotsakis, "Infation and the conformal structure of higher-order gravity
theories", Physics Letters B, 214-515(518), Dec. 1988.

\bibitem{23} T. Jacobson, D. Mattingly, 2001, "Gravity with a dynamical preferred frame", Phys. Rev. D64, 024028.

\bibitem{24} E. W. Kolb, M. S. Turner, "The Early Universe", Addison-Wesley, California, 1992.

\bibitem{25} Y. Fujii, K.-I. Maeda, "The Scalar–Tensor Theory of Gravitation", Cambridge Monographs
onMathematical Physics, (Cambridge University Press, Cambridge; New York, 2003).

\bibitem{26} G. Magnano, L.M. Sokolowski, "Physical equivalence between nonlinear gravity theories
and a general-relativistic self-gravitating scalar field", Phys. Rev. D, 50, 5039–5059, (1994).

\bibitem{27} J.L. Synge, "Relativity: the General Theory", North Holland, Amsterdam, 1955.

\bibitem{28} N.D. Birrell, P.C. and Davies, "Quantum Fields in Curved Space", Cambridge University
Press, Cambridge, 1982.

\bibitem{29} A. Linde, "Particle Physics and Inflationary Cosmology", Harwood Academic Publishers, Switzerland, 1990.

\bibitem{30} J. W. York, "Role of conformal three geometry in the dynamics of
gravitation", Phys. Rev. Lett. 28 (1972) 1082.

\bibitem{31} M. S. Madsen and J. D. Barrow, "De Sitter Ground States And
Boundary Terms In Generalized Gravity", Nucl. Phys. B 323 (1989)242.

\bibitem{32} R. Maartens, "Brane world gravity", 2004, Living Rev. Rel. 7, 7.

\bibitem{33} G. O. Abell, "The distribution of rich clusters of galaxies", Ap. J., Suppl. 3, 211 (1958).

\bibitem{34} C. W. Misner, K. S. Thorne, and J. A. Wheeler, "Gravitation", Freeman Press, San Francisco, 1973.

\bibitem{35} C. W. Misner, "The Isotropy of the Universe" 1968, Astrophys. J. 151, 431.

\bibitem{36} V. F. Mukhanov,"CMB, Quantum Fluctuations and the Predictive Power of Inflation", 2003, eprint astro-ph/0303077.

\bibitem{37} V. F. Mukhanov and R. H. Brandenberger, "A Nonsingular universe", Phys.Rev.Lett. 68 (1992) 1969-1972.

\bibitem{38} N. H. Barth, "The Fourth Order Gravitational Action For Manifolds
With Boundaries", Class. Quant. Grav. 2, 497 (1985).

\bibitem{39} S. Perlmutter, et al. (Supernova Cosmology Project), 1998, Nature 391, 51.

\bibitem{40} A. G. Riess, et al. (Supernova Search Team), 2004, Astro-phys. J. 607, 665.

\bibitem{41} H.-J. Schmidt, "Fourth order gravity: Equations, history, and applications to cosmology", 2007, Int. J. Geom. Meth. Phys. 4, 209.

\bibitem{42} T. P. Sotiriou, "f(R) gravity and scalar-tensor theory", 2006, Class. Quant. Grav. 23, 5117.

\bibitem{43} T. P. Sotiriou, "Modified Actions for Gravity: Theory and Phenomenology "-PhD Thesis ( arXiv:0710.4438v1 [gr-qc])

\bibitem{44} T. P. Sotiriou, S. Liberati, "The Metric-affine formalism of f(R) gravity", 2007, J. Phys. Conf. Ser. 68, 012022. ;

\bibitem{45} T. P. Sotiriou, S. Liberati, "Metric-affine f(R) theories of gravity " 2007, Annals Phys. 322, 935.

\bibitem{46} A. Balcerzak and M. P. Dabrowski, "Gibbons-Hawking Boundary Terms and Junction Conditions for Higher-Order Brane Gravity Models", arXiv:0804.0855 [hep-th].

\bibitem{47} R. Casadio and A. Gruppuso, "On boundary terms and conformal transformations in curved spacetimes", Int. J. Mod. Phys. D 11, 703(2002)

\bibitem{48} S. Nojiri and S. D. Odintsov, "Finite gravitational action for higher derivative and stringy gravities", Phys. Rev. D 62, 064018 (2000)

\bibitem{49} E. Dyer and K. Hinterbichler. "Boundary Terms, Variational Principles and Higher Derivative Modified
Gravity". Phys. Rev. D. 79 (2009):024028.

\bibitem{50} M. Ferraris, M. Francaviglia, I. Volovich, "The Universality of Einstein Equations", 1992, eprint gr-qc/9303007.

\bibitem{51} A. Guarnizo, L. Castaeda, and J. M. Tejeiro. "Boundary term in metric f(R) gravity: field
equations in the metric formalism". Gen. Rel. Grav. 42:2713, 2010, 1002.0617.

\bibitem{52} T. P. Sotiriou and V. Faraoni. "f(R) theories of gravity". Reviews of Modern Physics, 82:451(497),
Jan. 2010, 0805.1726.

\bibitem{53} S. Weinberg, "The cosmological constant problem",  1989, Rev. Mod. Phys. 61, 1.

\bibitem{54} S. Capozziello, S. Carloni, and A. Troisi. "Quintessence without scalar fields". ArXiv Astrophysics
e-prints, Mar. 2003, arXiv:astro-ph 0303041.

\bibitem{55} S. M. Carroll, V. Duvvuri, M. Trodden, and M. S. Turner. "Is cosmic speed-up due to new
gravitational physics?", Phys. Rev. D, 70(4):043528

\bibitem{56} A. A. Starobinsky, "A new type of isotropic cosmological models without singularity", Physics
Letters B, 91(1):99-102, 1980.

\bibitem{57} S. Capozziello, "Curvature quintessence", Int. J. Mod. Phys. D11, 483 (2002).

\bibitem{58} F. Zwicky, K. Rudnicki, Ap. J., 137, 707 (1963); Z. Astrofiz., 64, 246 (1966).

\bibitem{59} A.A. Penzias, R.W. Wilson, "A Measurement of Excess Antenna Temperature at 4080 Mc/s.", Ap. J. 142, 419 (1965).

\bibitem{60} C. V. I. Charlier, Arkiv. Mat. Astr. Fys., 4, No. 24 (1908); ibid., 16, No. 22 (1922).

\bibitem{61} II. Bondi, T. Gold, "The Steady-State Theory of the Expanding Universe", Mon. Rot. Roy. Astron. Soc., 108, 252 (1948).

\bibitem{62} F. Hoyle, " A New Model for the Expanding Universe ", Mon. Rot. Roy. Astron. Soc., 108, 372 (1948); ibid., 109, 365 (1949).

\bibitem{63} L. Pieronero, "The Fractal Structure of the Universe: Correlations of Galaxies and
Clusters". Physica A (144): 257 (1987).

\bibitem{64} S. Carroll, Spacetime and Geometry: "An introduction to General Relativity", Addison Wesley, 2004.

\bibitem{65} L.D. Landau, E.M. Lifshitz, "The Classical Theory of Fields. Vol. 2 (3rd ed.)", Pergamon Press, 1971.

\bibitem{66} T. Padmanabhan, "Structure Formation in the Universe", Cambridge University Press, 1993.

\bibitem{67} A. D. Dolgov, M. Kawasaki, "Can modified gravity explain accelerated cosmic expansion?", Phys. lett. B 573, 1 (2003).

\bibitem{68} J. D. Evans, Lisa. M. H. Hall, P. Caillol , "Standard cosmological evolution in a wide range of f(R) models", Phys. Rev. D 77, 083514 (2008).

\bibitem{69} L. Amendola, R. Gannouji, D. Polarski, S. Tsujikawa, "Conditions for the cosmological viability of dark energy models", Phys. Rev. D 75, 083504 (2007);
     L. Amendola, D. Polarski, S. Tsujikawa, "Are f(R) dark energy models cosmologically viable ?", Phys. Rev. Lett98, 131302 (2007).

\bibitem{70} J.D. Barrow, "The premature recollapse problem in closed inflationary universes", Nucl.
Phys. B, 296, 697–709, (1988).

\bibitem{71} J.D. Barrow, S. Cotsakis, "Inflation and the Conformal Structure of Higher-Order
Gravity Theories", Phys. Lett. B, 214, 515–518, (1988)

\bibitem{72} J.N. Islam, "An Introduction to Mathematical Cosmology", Cambridge University Press, 1992.

\bibitem{73} S. Gottloeber, H. J. Haubold, J.P. Muecket, V. Mueller, "Early evolution of the universe and formation of structure", Berlin, Akademie-Verlag Berlin, 1990.

\bibitem{74} M. Abramowitz,  I. A. Stegun, "Handbook of Mathematical Physics", Dover, 1972.

\bibitem{75} L. Amendola, D. Polarski, S. Tsujikawa, "Are f(R) dark energy models cosmologically
viable?", Phys. Rev. Lett., 98, 131302, (2007)

\bibitem{76} H.A. Buchdahl, "Non-linear Lagrangians and cosmological theory", Mon. Not. Roy. Astr. Soc. {\bf 150}, 1 (1970)
    
\bibitem{77} S. Nojiri, S.D. Odintsov, "Introduction to Modified Gravity and Gravitational Alternative for Dark Energy", ECONF C0602061:06, 2006; Int.J.Geom.Meth.Mod.Phys.4:115-146, 2007.
    
\bibitem{78} S. Nojiri, S.D. Odintsov ,"Unified cosmic history in modified gravity: from F(R) theory to Lorentz non-invariant models", Phys.Rept.505:59-144, 2011.
    
\bibitem{79} S. Nojiri, S.D. Odintsov, "Modified gravity with negative and positive powers of the curvature: unification of the inflation and of the cosmic acceleration", Phys.Rev.D68:123512, 2003.


\end{thebibliography}
\end{document}